\documentclass[fleqn,usenatbib]{mnras}

\usepackage{newtxtext,newtxmath}
\usepackage{lineno}
\usepackage{multirow}

\usepackage[T1]{fontenc}

\newcommand{\hii}{H {\sc i} }
\newcommand{\civ}{C {\sc iv}}
\newcommand{\heii}{He {\sc ii}}

\DeclareRobustCommand{\VAN}[3]{#2}
\let\VANthebibliography\thebibliography
\def\thebibliography{\DeclareRobustCommand{\VAN}[3]{##3}\VANthebibliography}

\usepackage{graphicx}	
\usepackage{amsmath}	

\title[Zhang, S. et al. 2024]{Testing the AGN unified model with simulated emission lines from the circumgalactic medium (CGM)}

\author[S. Zhang et al.]{
Shiwu Zhang$^{1}$,\thanks{E-mail: zsw@zhejianglab.com}
Zheng Cai$^{2}$, 
Aura Obreja$^{3}$, 
Fabrizio Arrigoni Battaia$^{4}$, 
L{\'e}o Michel-Dansac$^{5}$,
\newauthor
J{\'e}r{\'e}my Blaizot$^{6}$,
Donghui Quan$^{1}$, 
Mingyu Li$^{2}$
\\
$^{1}$Research Center for Astronomical Computing, Zhejiang Laboratory, Hangzhou 311100, China\\
$^{2}$Department of Astronomy, Tsinghua University, Beijing 100084, China\\
$^{3}$ Interdisziplin\"ares Zentrum f\"ur Wissenschaftliches Rechnen, Universit\"at Heidelberg, Im Neuenheimer Feld 205, D-69120 Heidelberg, Germany\\ 
$^{4}$ Max-Planck-Institut f$\ddot{u}$r Astrophysik, Karl-Schwarzschild-Stra$\beta$e 1, D-85748 Garching bei M$\ddot{u}$nchen, Germany\\
$^{5}$ Aix Marseille Univ., CNRS, CNES, Laboratoire d’Astrophysique de Marseille, Marseille, France\\
$^{6}$ Centre de Recherche Astrophysique de Lyon UMR5574, Univ Lyon, Univ Lyon 1, ENS de Lyon, CNRS, F-69230 Saint-Genis-Laval, France\\
}

\begin{document}
\label{firstpage}
\pagerange{\pageref{firstpage}--\pageref{lastpage}}
\maketitle


\begin{abstract}
The CGM around unobscured AGN has received much attention in recent years. 
Comparatively, nebulae associated with obscured AGN are less studied. 
Here, we simulate the Ly$\alpha$, H$\alpha$, and \heii \ nebulae around the two types of AGN at $z=2-3$ with ten massive systems from the FIRE simulations based on the unified model to show their differences and to test if they can be used to constrain the AGN model. 
We post-process the data with the \textsc{cloudy} and the Ly$\alpha$ radiative transfer code, \textsc{rascas}. 
Overall, we find that the Ly$\alpha$ nebulae around the unobscured AGN (type-I nebulae) and obscured AGN (type-II nebulae) do not exhibit significant differences in the luminosity, area, and \heii/Ly$\alpha$ when the simulated cutout is set to the halo virial radius. 
Whereas, the type-II nebulae exhibit less symmetric morphologies, flatter surface brightness profiles, and larger emission line widths (at $R\geq 10$ kpc) than those of the type-I nebulae. 
These nebulae properties exhibit complicated correlations with the AGN, indicating that nebulae observations can be applied to constrain the AGN engine. 
However, independent observations on nebulae in the mentioned emissions are insufficient to test the unified model as a priori in observations is not possible to know the direction and opening angle of the ionization cone. 
We prompt that the joint observations of Ly$\alpha$ nebulae and radio jets can help to reveal the ionization cone to probe the unified model. 
Our calculations suggest that this method requires $\geq 75$ type-II Ly$\alpha$ nebulae with current instruments to reach a confidence level of $\geq 95\%$.

\end{abstract}



\section{Introduction}
Galaxy evolution is highly regulated by the circumgalactic medium (CGM) which is loosely defined as the gas within the virial radius of the host halo and beyond the stellar disk \citep{Tumlinson2017,Faucher2023}. 
Since state-of-art integral field spectrographs (IFS) such as the Multi Unit Spectroscopic Explorer (MUSE) \citep{Bacon2010} on the Very Large Telescope (VLT) and the Keck Cosmic Web Imager (KCWI) \citep{Morrissey2018} on the Keck Telescope have been available, routinely detected extended emission regions, namely nebulae, provides us a more direct way compared to absorption \citep{Hennawi2006,Prochaska2013,Hennawi2013,Prochaska2014,Rudie2019,Faucher2023} to reveal the CGM properties \citep{Fumagalli2024}. 
AGN with typical luminosities of $L_{\rm AGN}=10^{45 - 48}$ erg s$^{-1}$ \citep{Padovani2017,Fawcett2023} are one of the best targets for detecting this emission.

In observations, depending on whether the ultraviolet (UV), X-ray continuum, and broad emission lines of the AGN are obscured, AGN is divided into unobscured (also known as the type-I AGN) and obscured (also known as the type-II AGN) type \citep{Netzer2015}. 
For the unobscured AGN, many IFS and narrowband (NB) surveys \citep{Heckman1991,Weidinger2004,Weidinger2005,Christensen2006,Borisova2016,Fab2016,Fab2019,Cai2019,Farina2019,OSullivan2020,Mackenzie2021,Fossati2021,Herwig2024,Li2024} have been conducted to search for the nebulae around them. 
These surveys discover $\geq 300$ Ly$\alpha$ nebulae at $z=2-6$, probing the CGM from a few tens to hundreds of kiloparsecs around the targeted quasars. 
Recently, simulations on nebulae around the unobscured AGN at $z\geq 3$ were also performed to uncover the connection between the AGN and nebulae, highlighting the importance of AGN feedback \citep{Costa2022} and of the AGN ionization cone \citep{Obreja2024} in matching the observed emissions. 
Whereas, observations on the nebulae associated with the obscured AGN are limited to only a few cases \citep{Cai2017,Law2018,Brok2020,Zhang2023a,Zhang2023b}. 
No simulation on the nebulae around the obscured AGN has been conducted yet.  
In this work, we focus on simulating the nebulae around both types of AGN to fill in the blanks.


An essential application of such simulations is to explore the potential differences between the nebulae around the two types of AGNs, which can be applied to test the AGN unified model. 
This model postulates that the supermassive black hole (SMBH), accretion disc, hot corona, and the broad line region (BLR) are surrounded by an optically thick dusty torus with the hydrogen column density of $N_{\rm H}\geq 10^{22}$ cm$^{-2}$ \citep{Antonucci1993,Urry1995}. 
Whether the AGN is obscured depends on whether the sightline passes through the torus. 
In the local Universe, this model has been probed by direct imaging on tori on a physical scale of $1- 100$ pc \citep{Tristram2014,Carilli2019}. 
At the high redshift ($z\geq 2$), only a few observations \citep{Prochaska2013,Johnson2015,Wang2024} show indirect evidence of the unified model by revealing the anisotropic AGN radiation, i.e. the ionization cone.




Whereas, recent observations raise tension on the unified model at high redshift. 
The James Webb Space Telescope ({\it JWST}) unveils a new population of faint obscured AGN with a number density of $\approx 10^{-5}$ cMpc$^{-1}$ at $z\geq 4$ which is orders of magnitude higher than the quasar density at the same redshift \citep{Matthee2024,Maiolino2023,Harikane2023,Greene2024,Kocevski2024,Kokorev2024}. 
The high density of the obscured AGN is inconsistent with the unified model \citep{Hopkins2012,Roth2012,Hopkins2016} which predicts comparable fractions for the two types of AGN \citep{Hopkins2012,Roth2012,Hopkins2016}. 
This is also inconsistent with previous X-ray observations which show obscured fraction ranges of $0.2 - 0.8$ \citep{Hasinger2008,Buchner2015,Vijarnwannaluk2022}. 
The connection between the SMBH mass and the broad line strength revealed in these {\it JWST} observations favor the evolutionary model where, as the SMBH grows up, the AGN transfers from the obscured phase to the unobscured phase by expelling its surrounding materials in the interstellar medium (ISM) or the circumnuclear region through the AGN feedback \citep{Sanders1988,Granato2004,Hopkins2008,Alexander2012}.

These new observations indicate that the picture of the high-$z$ AGN remains quite vague. 
Whereas, uncovering this picture by directly imaging the AGN inner structure is difficult at high redshift. 
Resolving the torus with physical scales of $\leq 100$ pc requires an angular resolution of $\approx 0.01$ arcsec at $\approx 30 \ \mu$m (by adopting the results of \cite{Tristram2014} and \cite{Carilli2019} in the local Universe). 
This is beyond the capability of current state-of-art facilities like the {\it JWST} \citep{Dicken2024}. 
Although the warm dust with a scale of $\approx 200$ pc of an AGN at $z\approx 7$ has been imaged by recent observations with the Atacama Large Millimeter Array (ALMA) \citep{Meyer2025}, the torus is still not resolved. 
Since the ionization cone which is a key prediction of the unified model imprints on the CGM nebulae, the nebulae can be used to test this model by uncovering the cone. 
\cite{Brok2020} uses the VLT/MUSE to observe Ly$\alpha$ nebulae around the two types of nebulae at $z\approx 3$ and find tentative evidence supporting the unified model. 
By simulating the Ly$\alpha$ nebulae based on the zoom-in simulations, \cite{Obreja2024} further showcases that some of the nebulae observables can be used to pinpoint properties of the AGN. 
These works suggest that the CGM nebulae could be useful for constraining the AGN engine.



One of the highest resolution cosmological simulations of massive galaxies is the FIRE simulations \citep{Hopkins2014}. 
The fact that the ISM and CGM of these simulations are by construction multi-phase makes them ideal for our experiment of simulating nebulae. 
Specifically, four massive systems with the halo mass of $10^{11.1  - 12.5} \ M_{\odot}$ at $z=2-3$ \citep{Angle2017} selected from the A-series of the FIRE-1 MassiveFIRE suite \citep{Feldmann2016,Feldmann2017} are used in this work. 
In Sec.~\ref{simulations}, we describe the cosmological simulations, how the spectral energy distribution (SED) of the ionizing sources are constructed, how the photoionization is modeled, how the radiative transfer (RT) calculations are performed for the Ly$\alpha$ emission, and how the mock observables are generated. 
In Sec.~\ref{sim_results}, we compare the morphology, surface brightness, and spectral profile of the nebulae around the two types of AGN. 
In Sec.~\ref{sim_discussion}, we discuss the connection between the nebulae and the AGN and how the nebulae can be used to constrain the AGN engine at the high redshift. 
In Sec.~\ref{Conclusions}, we summarize our main conclusions. 








\section{Simulations} \label{simulations}

In this work, we use the public data \footnote{\url{https://flathub.flatironinstitute.org/fire}} of massive systems at $z=2-3$ in the suite of FIRE cosmological zoom-in simulations \citep{Angle2017}. 
The FIRE simulations are run with the \textsc{gizmo} code \footnote{\url{http://www.tapir.caltech.edu/~phopkins/Site/GIZMO.html}} in the mode of pressure-entropy implementation of smooth particle hydrodynamics (SPH). 
The gravitational force is calculated by \textsc{gadget-3} \citep{Springel2005}. 
These simulations include stellar feedback effects \citep{Hopkins2014}, and follow the growth of SMBHs, but do not consider any AGN feedback on the gas. 

The star particles represent the single stellar populations that are converted from the molecular gas particles with the hydrogen number density $n_{\rm H}\geq 5 - 50$ cm$^{-3}$ \citep{Angle2017}. 
The emerging Kennicutt-Schmidt (KS) relation is not sensitive to this density threshold in FIRE simulations. 
Once a star particle forms, stellar feedback is implemented as stellar radiation pressure, \hii \ photoionization and photoelectric heating, Type I and Type II supernovae (SNe), and stellar winds \citep{Leitherer1999}. 
The gas is then metal-enriched by the materials ejected from SNe and stellar winds. 
The radiative cooling from the primordial gas \citep{Katz1996} is included in the presence of a uniform photoionizing background \citep{Faucher2009}. 
The metal-line cooling rates on an element-by-element basis \citep{Wiersma2009} and molecular cooling processes are also taken into account. 

\cite{Angle2017} selected the zoom-in regions from a large volume by targeting halos in the mass range of M$_{\rm h}\approx 10^{10 - 13}$ M$_{\odot}$ at $z=0$.  
These regions yield the baryonic particle mass and minimum force softening length of $m_{b}\leq  3.7\times 10^{5}$ M$_{\odot}$ and $\epsilon_{b}\leq 21$ pc, respectively. 
Please refer to \cite{Angle2017} for detailed information. 
The force softening length in their simulations is identical to the gas smoothing length which encloses $\approx 60$ SPH neighbors.  
These resolutions allow us to trace the gas particles with the density of $\geq 10^{3}$ cm$^{-3}$. 
They also adopt the standard flat $\Lambda$ cold dark matter (CDM) cosmology with $h\approx 0.7$, $\Omega_{\rm M}=1-\Omega_{\Lambda}\approx0.27$, and $\Omega_{\rm b}\approx0.046$ \citep{Planck_Collaboration2014}. 

 \subsection{Massive system selection} \label{halo_selection}
\begin{figure}
    \centering
    \includegraphics[width=\columnwidth]{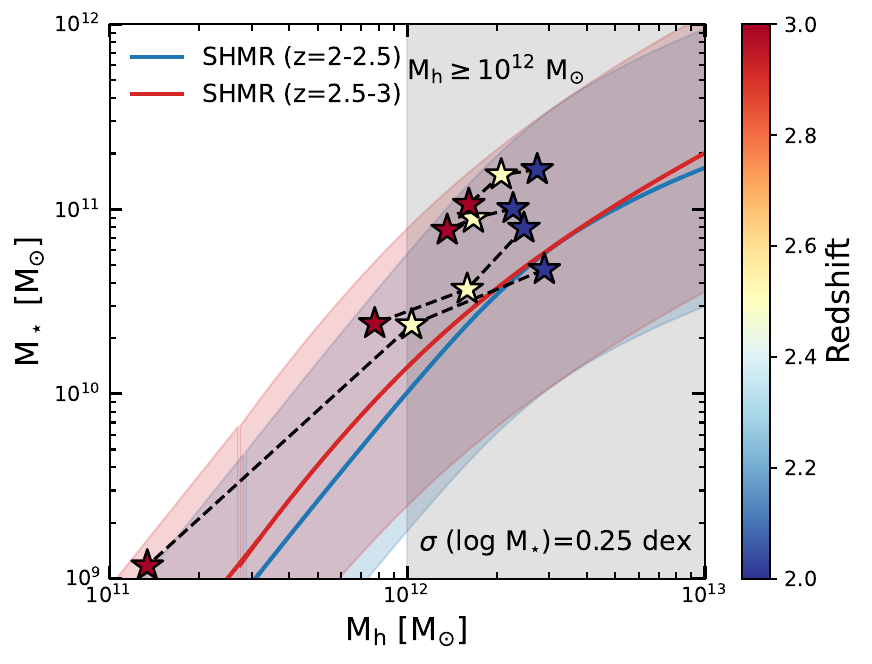}
    \caption{The halo mass versus the stellar mass of 12 massive systems in FIRE simulations and the observed SHMR (dotted-dashed lines) at $z=2-3$ \citep{Shuntov2022}.  
    The color of the stars denotes the redshift of the halos. 
    The red and blue lines represent the SHMR in the redshift ranges of $z=2-2.5$ and $z=2.5-3$, respectively, with the shadows denoting the 3-$\sigma$ stellar mass scatter where $\sigma$ (log M$_{\star}$)=0.25 \citep{Matthee2017}. 
    The gray shadow denotes the halo mass range of M$_{\rm h}\geq10^{12}$ M$_{\odot}$. 
    The dashed line connects the same halo at different snapshots. 
    Given the selection criteria in Sec.~\ref{halo_selection}, two halos at $z=3.0$, four halos at $z=2.5$, and four halos at $z=2.0$ are selected.}
    \label{SHMR}
\end{figure}

Only four massive systems are published from the FIRE simulations. 
Since there are only three snapshots of $z=2.0$, $z=2.5$, and $z=3.0$ in $z=2-3$, we have 12 massive systems in total. 
Previous observations demonstrate that the AGN usually inhabits halo with M$_{\rm h}\geq 10^{12}$ M$_{\odot}$ at $z=2-3$ \citep{Geach2019,Aird2021}. 
In this work, only massive systems with M$_{\rm h}\geq 10^{12}$ M$_{\odot}$ and consistent with the stellar-mass-to-halo mass relation (SHMR) under the 3-$\sigma$ scatter are selected. 
Fig.~\ref{SHMR} presents the halo mass versus the stellar mass of the 12 massive systems and the observed SHMR at $z=2-3$ \citep{Shuntov2022}. 
Given the above halo mass threshold and the stellar mass scatter of $\sigma$(log M$_{\star}$)=0.25 dex at $z=2-3$ \citep{Matthee2017}, 10 out of the 12 halos are selected (Tab.~\ref{halo_info}). 
The halo mass, stellar mass, and gas mass (within the virial radius) are in the range of M$_{\rm h}=10^{12.01 - 12.46}$ M$_{\odot}$, M$_{\star}=10^{10.37 - 11.21}$ M$_{\odot}$, and M$_{\rm gas}=10^{10.91 - 11.57}$ M$_{\odot}$, respectively. 
This implies an integrated baryon conversion efficiency of $\epsilon=M_{\star}/(M_{\star}+M_{\rm gas})=0.10 - 0.51$, which is within the 3-$\sigma$ scatter of the abundance matching predictions for halos at $z=2-3$ where $\epsilon(M_{\rm h}=10^{12} \ M_{\odot}, z=2)=0.18^{+0.33}_{-0.15}$ \citep{Moster2018}. 
Since the $M_{\rm BH}$ of the selected halos is not within the measured $M_{\star}-M_{\rm BH}$ relation at $z=2$ \citep{Zhang2023_mstar_mbh}, we adopt this measured relation to estimate the $M_{\rm BH}$ from the stellar mass. 
This relation yields the $M_{\rm BH}=10^{8.25-9.07} M_{\odot}$ (Tab.~\ref{halo_info}). 

\begin{table*}
    \setlength{\tabcolsep}{1.5pt}
    \centering
    \begin{tabular}{ccccccccccc}
    \hline
    \hline
         & A$_{\rm  1}$ ($z=2.0$)  & A$_{\rm  1}$ ($z=2.5$) & A$_{\rm  1}$ ($z=3.0$) & A$_{\rm  2}$ ($z=2.0$) & A$_{\rm  2}$ ($z=2.5$) & A$_{\rm  2}$ ($z=3.0$) & A$_{\rm  4}$ ($z=2.0$) & A$_{\rm 4}$ ($z=2.5$)& A$_{\rm  8}$ ($z=2.0$) & A$_{\rm 8}$ ($z=2.5$)\\
         \hline
       M$_{\rm h}$ [M$_{\odot}$] & $10^{12.36}$ & $10^{12.22}$ & $10^{12.13}$ & $10^{12.44}$ &$10^{12.32}$ &$10^{12.21}$ &$10^{12.39}$ &$10^{12.20}$ &$10^{12.46}$ &$10^{12.01}$ \\
        M$_{\star}$ [M$_{\odot}$] &$10^{11.00}$  &$10^{10.95}$  &$10^{10.89}$  &$10^{11.21}$  &$10^{11.19}$ &$10^{11.03}$ &$10^{10.90}$ &$10^{10.57}$ &$10^{10.67}$ &$10^{10.37}$  \\
        M$_{\rm BH}$ [M$_{\odot}$] & $10^{8.86}$ & $10^{8.81}$ & $10^{8.75}$ & $10^{9.07}$ & $10^{9.04}$ & $10^{8.89}$& $10^{8.75}$&$10^{8.44}$ & $10^{8.54}$& $10^{8.25}$ \\
        M$_{\rm gas}$ [M$_{\odot}$] &$10^{11.23}$  & $10^{11.02}$ &  $10^{10.91}$& $10^{11.38}$ &$10^{11.17}$ &$10^{11.06}$ &$10^{11.36}$ &$10^{11.19}$ &$10^{11.57}$ &$10^{11.07}$  \\
        $\epsilon$ & $0.37$ &$0.46$  &$0.49$  &$0.41$  &$0.51$ &$0.48$ &$0.26$ &$0.19$ &$0.11$ &$0.17$  \\
    \hline
    \hline
    \end{tabular}
    \caption{{\bf The properties of massive systems selected from FIRE simulations.} Each row from the top to the bottom represents the halo mass, stellar mass, SMBH mass derived by the stellar mass with the $M_{\star}-M_{\rm BH}$ relation at $z\approx2$ \citep{Zhang2023_mstar_mbh}, and the integrated baryon conversion efficiency. 
    Since these halos are within the 3-$\sigma$ scatter of the SHMR at $z=2-3$ and have the halo mass of $\geq 10^{12}$ M$_{\odot}$, they are selected for simulating the nebulae.}
    \label{halo_info}
\end{table*}

\subsection{Construction of the mock observations} \label{observations_mock}
\begin{figure*}
    \centering
    \includegraphics[width=\textwidth]{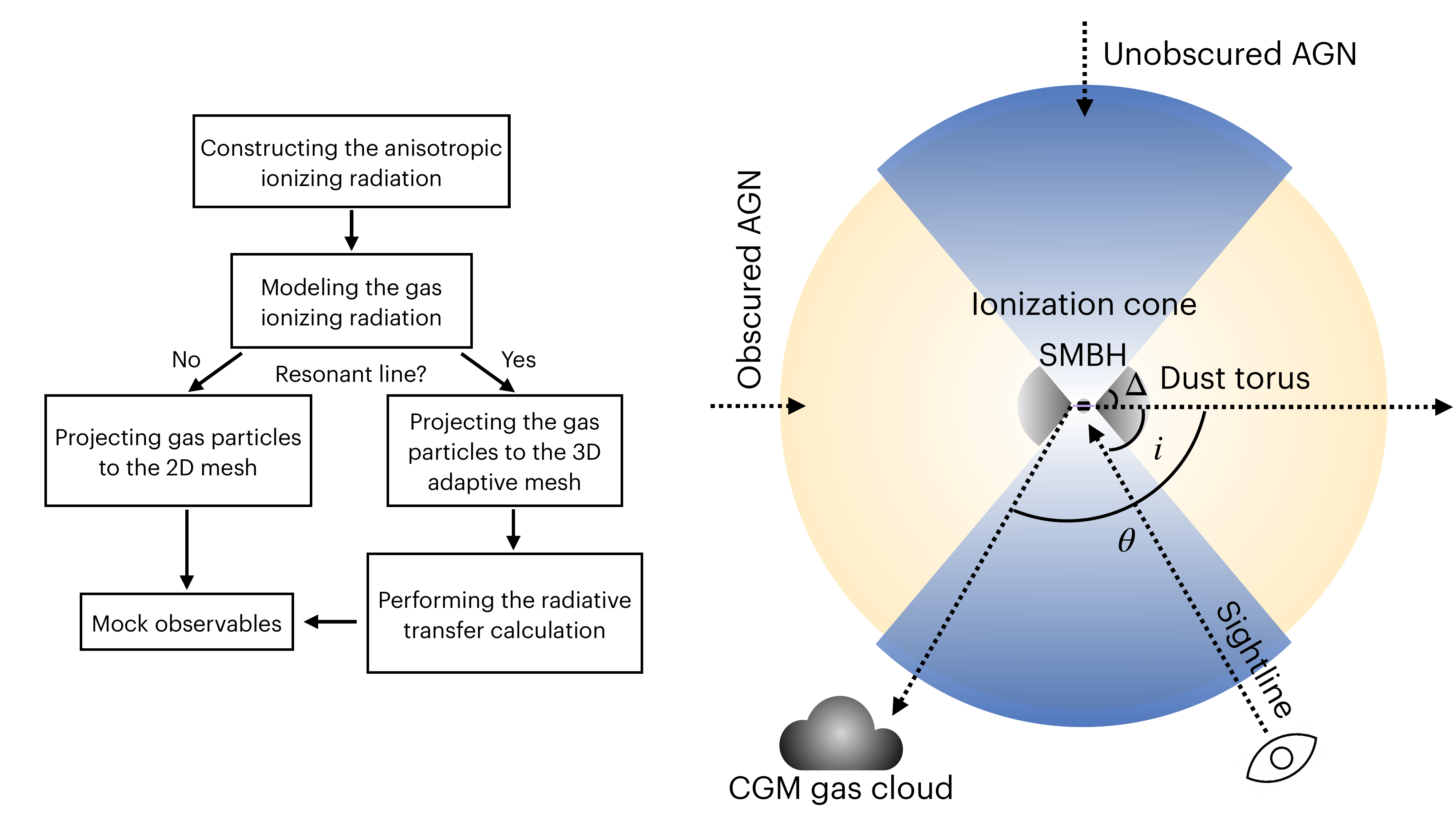}
    \caption{{\bf Left:} The workflow for producing the mock observables. 
    The anisotropic AGN radiation is constructed with the  \textsc{x-cigale} \citep{Boquien2019,Yang2020} code. 
    The gas emissivities are then calculated by running \textsc{cloudy} \citep{Ferland2017} on the grid of parameters. 
    These emissivities are assigned to the gas particles by cross-matching the parameters with the gas properties. 
    For the Ly$\alpha$ without the processing of \textsc{rascas}, H$\alpha$, and \civ, the gas particles are projected to the 2D mesh to produce the mock images. 
    To account for the resonant scattering of Ly$\alpha$ photons in the CGM, we employ \textsc{rascas} code \citep{Michel2020} to do the RT calculation after projecting the gas particles to a 3D adaptive mesh.
    {\bf Right:} The sketch of the model geometry. 
    We let the half-opening angle ($\Delta$) of the dusty torus and the inclination ($\theta$) of the gas cloud ranges in $50^{\rm o}-70^{\rm o}$ and $0^{\rm o} - 360^{\rm o}$ in steps of $10^{\rm o}$. 
    For each placement of the ionization cone, we randomly select a viewing angle ($i$) in and out of the cone to produce the mock observables for the two types of AGN.}
    \label{workflow_geometry}
\end{figure*}

Fig.~\ref{workflow_geometry} shows the workflow and geometry of producing the mock observables for the Ly$\alpha$, H$\alpha$, and \heii \ emissions. 
The details of each step shown in this figure are presented in Appendix \ref{appendix_mockobs}. 
A brief summary is presented as follows:
\begin{itemize}
    \item {\bf Constructing the anisotropic ionizing radiation.} 
    The ionizing photons from the AGN, its host galaxy, and the ultraviolet background (UVB) are included in this study. 
    The SED of the UVB at $z=2-3$ is modeled by employing the results of \cite{Khaire2019}. 
    The \textsc{x-cigale} code \citep{Boquien2019,Yang2020} is employed to construct the spectral energy distribution (SED) of the AGN and its host galaxy. 
    Since the flux of the ionizing photons escaping from the host galaxy is orders of magnitude lower than those of the AGN, we simplify the modeling of the galaxy SED by setting it to be isotropic and only constraining its stellar mass, age, and metallicity. 
    
    A previous study shows that the opening angle of the AGN ionization cone should be mostly restricted to $40^{\rm o}\leq \alpha_{\rm cone}\leq 80^{\rm o}$ \citep{Fritz2006} corresponding to the half-opening angle of the dusty torus ($\Delta$) of $50^{\rm o}\leq \Delta \leq 70^{\rm o}$. 
    In this work, we let the $\Delta$ vary in this range in steps of $10^{\rm o}$. 
    Since the AGN unified model predicts that the AGN SED depends on the inclination ($\theta$), we let $\theta$ vary in $0^{\rm o} - 360^{\rm o}$ in steps of $10^{\rm o}$ where $\theta=90^{\rm o}, 270^{\rm o}$ denotes the perfect face-on condition and $\theta=0^{\rm o}, 180^{\rm o}$ denotes the perfect edge-on condition. 
    For each $\theta$, we apply the corresponding attenuation curve of dusty torus \citep{Stalevski2016} to the face-on SED ($\theta=90^{\rm o}$). 
    This ensures that the AGN SED with different inclinations share the same intrinsic properties. 
    For each simulated system, we select three random orientations to place the AGN ionization cone. 
    For each placement of the cone, we randomly select a viewing angle ($i$) in and out of the ionization cone to produce the mock image of nebulae for the two types of AGN.  
    These settings on the AGN anisotropic radiation yield $10\times 3\times 3\times 2=180$ mock observables for each emission line where the left-side numbers of the equation denote the number of systems, placements of the cone, settings of $\Delta$, and types of AGN, respectively.

    \item {\bf Modeling the gas ionizing radiation.} 
    We run the \textsc{cloudy} code \citep{Ferland2017} on a grid of parameters of hydrogen number density ($n_{\rm H}$), gas temperature ($T$), gas metallicity ($Z$), surface flux of the ionizing photon ($\Phi({\rm H})$), and the inclination ($\theta$) of gas particles to model the ionizing radiation of the CGM gas where $\theta$ comes from the SED modeling. 
    By cross-matching these parameter grids with the properties of the gas particles, emissivities of Ly$\alpha$, H$\alpha$, and \heii \ are assigned to the gas particles. 
    Note that we add this line on top of the AGN continuum and allow continuum pumping in the \textsc{cloudy} calculation to approximate the Ly$\alpha$ photons from the BLR.  
    The mock observables of H$\alpha$, \heii, and Ly$\alpha$ (without scattering for photons produced in situ) are constructed by projecting these post-processed gas particles to the two-dimensional (2D) mesh. 
    The 2D Gaussian profile with the full width of half maximum (FWHM) of $1''$ is convolved to the mesh to simulate the seeing. 
    \item {\bf Performing the radiative transfer (RT) calculation.} 
    For the process of Ly$\alpha$ resonantly scattering in the CGM, we employ the \textsc{rascas} code \citep{Michel2020} to do the radiative transfer (RT) calculation. 
    Since this code works on the adaptive mesh refinement (AMR) with the octree structure, we use the \textsc{yt-project}\footnote{\url{https://yt-project.org/}} which is a \textsc{python} module to project the gas particles to a three-dimensional (3D) adaptive mesh. 
    The mock observables are automatically generated by \textsc{rascas}. 
    Please see Appendix \ref{mock_diffuse_emission} for details.
\end{itemize}

\section{Results} \label{sim_results}

Although several works have been done to simulate the CGM nebulae around the unobscured AGN \citep{Byrohl2021,Costa2022,Obreja2024} at $z\geq 2$, the nebulae around obscured AGN have not been simulated yet. 
In the following text, the nebulae around the unobscured and obscured AGN are denoted as type-I and type-II nebulae, respectively. 
We focus on reproducing the two types of nebulae at $z=2-3$ based on the AGN unified model. 
The results allow us to explore the potential difference between the two types of nebulae, which could help us constrain the AGN model at these intermediate redshifts.


\subsection{Overview} \label{mock_image}
\begin{figure*}
    \centering
    \includegraphics[width=0.9\textwidth]{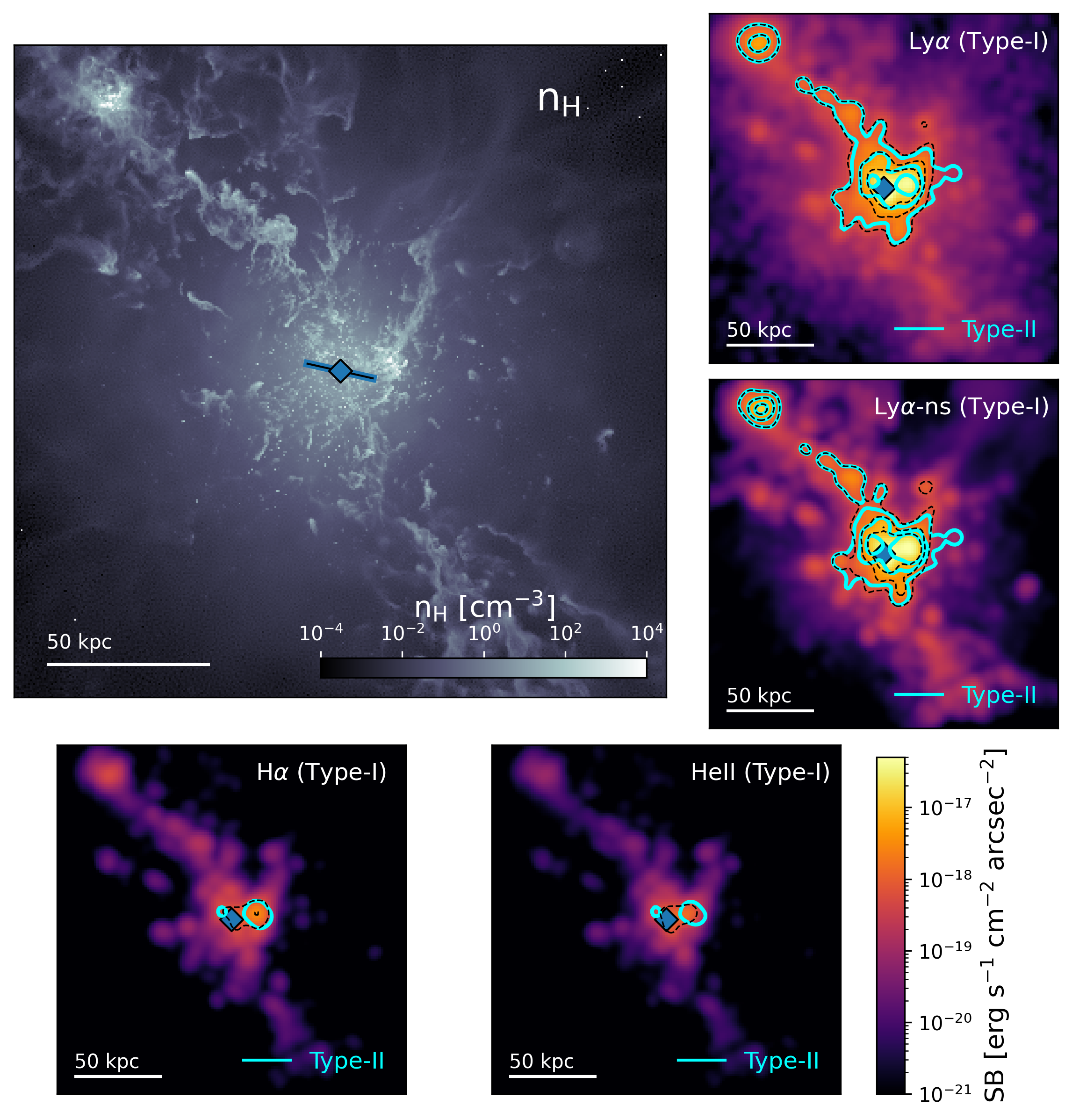}
    \caption{{\bf Upper left:} The hydrogen number density (n$_{\rm H}$) of the A$_{\rm 1}$ system at $z=2.0$ (Tab.~\ref{halo_info}) projected on the $x-y$ plane centered on the SMBH. 
    The physical width of this image is 200 kpc. 
    The blue diamond marks the position of the SMBH and the blue solid line denotes the orientation of the host galaxy projected on the $x-y$ plane. 
    {\bf Upper right two panels:} The mock images of the type-I Ly$\alpha$ nebulae with and without processing of \textsc{rascas}. 
    The type-II nebulae are overlaid on these images with cyan contours. 
    In the following figures, we denote the data of Ly$\alpha$ nebulae without processing of \textsc{rascas} as `Ly$\alpha$-ns'. 
    These images are smoothed with a Gaussian kernel of $1.0''$. 
    The contours denote [2$\sigma_{\rm SB}$, 10$\sigma_{\rm SB}$, 30$\sigma_{\rm SB}$] assuming $\sigma_{\rm SB}=3.0\times 10^{-19}$ erg s$^{-1}$ cm$^{-2}$ arcsec$^{-2}$. 
    Since the resonant scattering could bring the Ly$\alpha$ photons out to the large radius, the Ly$\alpha$ nebula with the processing of \textsc{rascas} is more extended than the nebula without the processing of \textsc{rascas}. 
    {\bf Bottom two panels:} Same as the upper right panels but for the H$\alpha$ and \heii \ nebulae. 
    These panels show that the two types of nebulae share a similar overall shape while acting differently in the central region because the projections of the ionization cone are different for the two sightlines.}
    \label{nH_vs_SB}
\end{figure*}

We begin in Fig.~\ref{nH_vs_SB} with the visual overview of the hydrogen number density and the mock images of the Ly$\alpha$, H$\alpha$, and \heii \ nebulae of the A$_{\rm 1}$ system at $z=2.0$ (Tab.~\ref{halo_info}). 
These mock images are generated by adopting $\Delta=60^{\rm o}$ which corresponds to the opening angle of the ionization cone of $60^{\rm o}$. 
We fix the sightline to the $z$-axis. 
The ionization cone is placed along the $z$-axis and $x$-axis to generate the mock images of the two types of nebulae. 
Since previous observations reach the seeing of $0.8'' - 2.1''$ and the 1-$\sigma$ SB limit of $ 2.5-46.0 \times 10^{-19}$ erg s$^{-1}$ cm$^{-2}$ arcsec$^{-2}$ \citep{Borisova2016,Cai2019,Fab2019,OSullivan2020,Li2024}, we adopt the values of FWHM$_{\rm seeing}=1''$ and $\sigma_{\rm SB}=3.0 \times 10^{-19}$ erg s$^{-1}$ cm$^{-2}$ arcsec$^{-2}$ in the following sections. 
Under the 2-$\sigma$ SB limit, Fig.~\ref{nH_vs_SB} shows that the Ly$\alpha$ nebula could extend to $\geq 150$ kpc and H$\alpha$, and \heii \ nebulae could extend to $\approx 20$ kpc. 
From this figure, we find that the two types of Ly$\alpha$ nebulae share a similar overall shape. 
An obvious difference is that the type-I Ly$\alpha$ nebula centers at a bright core while the type-II Ly$\alpha$ nebula has two cores in the central region. 
The H$\alpha$ and \heii \ nebulae exhibit the same trend. 
This is because the ionization cone projects in circular and bipolar shape respectively when the sightline is within and out of the cone. 
Besides, since the resonant scattering can shift photons from small to large radii, the Ly$\alpha$ nebula produced by \textsc{rascas} is more extended than the one without processing of \textsc{rascas}. 
Another interesting finding from Fig.~\ref{nH_vs_SB} is that the flux peak's positions of nebulae are inconsistent with their host galaxies. 
This is because the gas distribution is not centered on the host galaxy. 
Such offset has been already found in observations and simulations \citep{Fab2019,Drake2019,Claeyssens2022,Costa2022,Gonz2023}.

\begin{figure}
    \centering
    \includegraphics[width=\columnwidth]{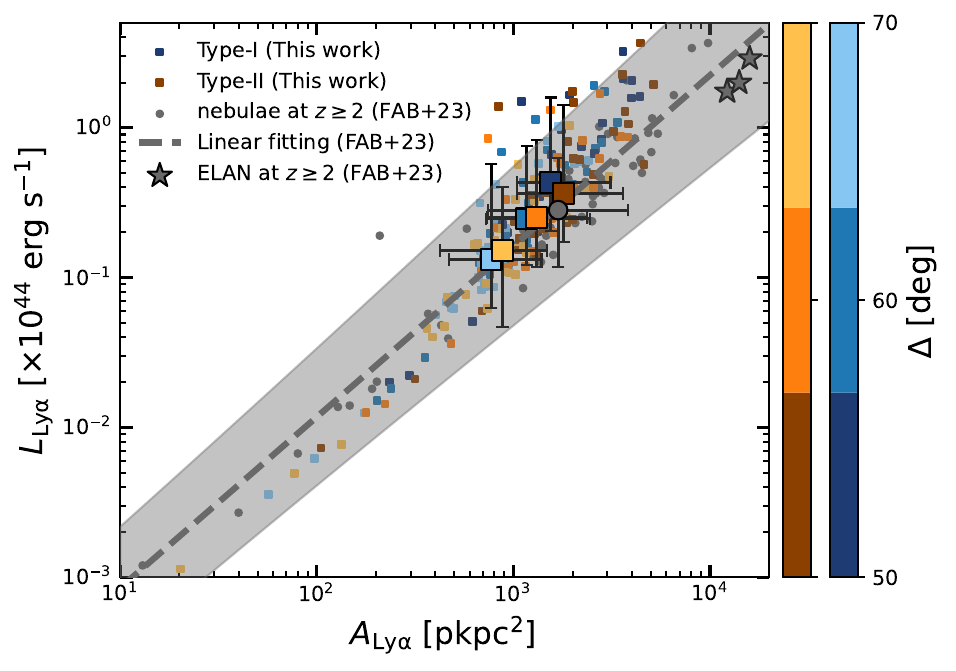}
    \caption{The luminosity-area relation of the Ly$\alpha$ nebulae at $z=2-3$. 
    The small blue and orange squares represent the simulated two types of nebulae, respectively. 
    The colormap denotes the half-opening angle of the torus ($\Delta$) which ranges in $50^{\rm o} - 70^{\rm o}$. 
    The gray dots, stars, and dashed line represent the observed Ly$\alpha$ nebulae \citep{Cai2019,Fab2019}, enormous Ly$\alpha$ nebulae (ELAN) \citep{Cantalupo2014,Hennawi2015,Fab2018}, and their best-fitting linear function. 
    The shadow denotes the 2-$\sigma$ uncertainty of the fitting. 
    The large squares and dots with errorbars represent the median values of simulated nebulae under different $\Delta$ and observed nebulae where the errorbar denotes the 16th and 84th percentiles. 
    The luminosity and area are calculated by employing the dimming-corrected SB threshold which corresponds to $\approx 3.5\times 10^{-18} \ {\rm erg \ s^{-1} \ cm^{-2} \ arcsec^{-2}}$ at $z\approx3.2$. 
    This figure suggests that our simulations could reproduce the observed luminosity-area relation with $\approx 95\%$ simulated Ly$\alpha$ nebulae in the shadow area of the best-fitting function.}
    \label{L_vs_D_lya}
\end{figure}

The luminosity-area relation of the 180 simulated Ly$\alpha$ nebulae is shown in Fig.~\ref{L_vs_D_lya}. 
The luminosity is calculated by summing up the flux within the dimming-corrected SB threshold corresponding to $\sigma_{\rm SB}\approx3.5\times 10^{-18} {\rm erg \ s^{-1} \ cm^{-2} \ arcsec^{-2}}$ at $z\approx3.2$ which is given by \cite{Fab2023} while the flux from the $1''\times 1''$ square centering on the AGN is excluded. 
The nebulae areas are also calculated with the same SB threshold. 
These calculations yield luminosities and areas of the Ly$\alpha$ nebulae ranging in $1.1\times 10^{41} - 3.7\times 10^{44} \ {\rm erg \ s^{-1}}$ and $20 - 5175 \ {\rm pkpc^{2}}$. 
The Welsch t-test returns the p-values of luminosity and area between the two types of nebulae to be $p_{\rm area}=45\%$ and $p_{\rm L}=65\%$. 
Given the p-value threshold of 5\%, these p-values indicate that the two types of Ly$\alpha$ nebulae do not show significant differences in luminosity and area. 

We also compare our simulated luminosity-area relation with the observed relation \citep{Fab2023} where the Ly$\alpha$ nebulae and enormous Ly$\alpha$ nebulae (ELAN) at $z=2-3$ are collected from previous observations \citep{Cantalupo2014,Hennawi2015,Fab2018,Cai2019,Fab2019}. 
Fig.~\ref{L_vs_D_lya} shows that $\approx 95\%$ simulated Ly$\alpha$ nebulae fall within the 2-$\sigma$ uncertainty (shadow area) of the best-fitting linear function (dashed line) to the observed nebulae, implying that our simulations could reproduce the observed luminosity-area relation. 
In particular, since our simulations cut the Ly$\alpha$ nebulae within the virial radius of the halo, the full luminosity-area relation can be retrieved by covering the faint part of this relation, as pointed out in \cite{Fab2023}. 
Moreover, the simulated Ly$\alpha$ nebulae move up along this relation as the $\Delta$ decreases (the opening angle of the ionization cone increases). 
This indicates that the opening angle of the cone could be one of the factors driving this relation if the unified model is correct. 
In the following sections, we present detailed comparisons between the two types of nebulae.

\subsection{Nebulae asymmetry} \label{nebulae_asymmetry}
The distribution of the cool gas and how it is illuminated determine the morphology of the nebulae. 
Although observations have shown that type-II Ly$\alpha$ nebulae are slightly asymmetric than the type-I nebulae at $z\geq 3$, which is recognized as the evidence supporting the unified model \citep{Brok2020}, there is no simulation exploring this scenario. 
We will explore if the two types of nebulae have different morphology in this section and discuss the connection between the morphology and AGN properties in Sec.~\ref{property_nebulae_AGN}. 

To quantify the symmetry of the nebula morphology, we adopt the same approach as \cite{Fab2019} and \cite{Cai2019} to calculate the flux-weighted dimensionless parameter, $\alpha_{\rm w}$ which is the ratio between the semiminor and semimajor axis of the nebulae. 
We specify the nebulae as the region above the 2-$\sigma$ SB limit where we adopt $\sigma_{\rm SB}=3.0\times 10^{-19}$ erg s$^{-1}$ cm$^{-2}$ arcsec$^{-2}$. 
We calculate the $\alpha_{\rm w}$ with Eq.~\ref{moment_for_alpha} \citep{Stoughton2002} 
\begin{subequations}
\label{moment_for_alpha}
    \begin{align}
        & M_{xx}=\langle \frac{(x-x_{\rm AGN})^{2}}{r^{2}} \rangle_{f} \\
        & M_{yy}=\langle \frac{(y-y_{\rm AGN})^{2}}{r^{2}} \rangle_{f} \\
        & M_{xy}=\langle \frac{(x-x_{\rm AGN})(y-y_{\rm AGN})}{r^{2}}\rangle_{f} \\ 
        & \alpha_{\rm w}=\frac{1-\sqrt{(M_{xx}-M_{yy})^{2}+(2M_{xy})^{2}}}{1+\sqrt{(M_{xx}-M_{yy})^{2}+(2M_{xy})^{2}}}
    \end{align}
\end{subequations}
where $M_{\rm xx}$, $M_{\rm yy}$, and $M_{\rm xy}$ are the second-order moments, ($x$, $y$) and ($x_{\rm AGN}$, $y_{\rm AGN}$) are the positions of a pixel and the central AGN (the position of the BH), and $r$ is the distance of a pixel to the AGN. 
This parameter ranges in $0-1$ with $\alpha_{\rm w}=1$ indicating that the nebula is perfectly circularly symmetric. 
We also calculate the $\alpha_{\rm uw}$ where the pixel flux is not taken as the weight. 
This parameter is adopted by \cite{Brok2020} and \cite{Wang2023} to avoid the domination of the central bright region. 
Since all pixels are weighted equally, $\alpha_{\rm uw}$ measures the symmetry of the overall shape of the nebulae. 
\begin{figure}
    \centering
    \includegraphics[width=\columnwidth]{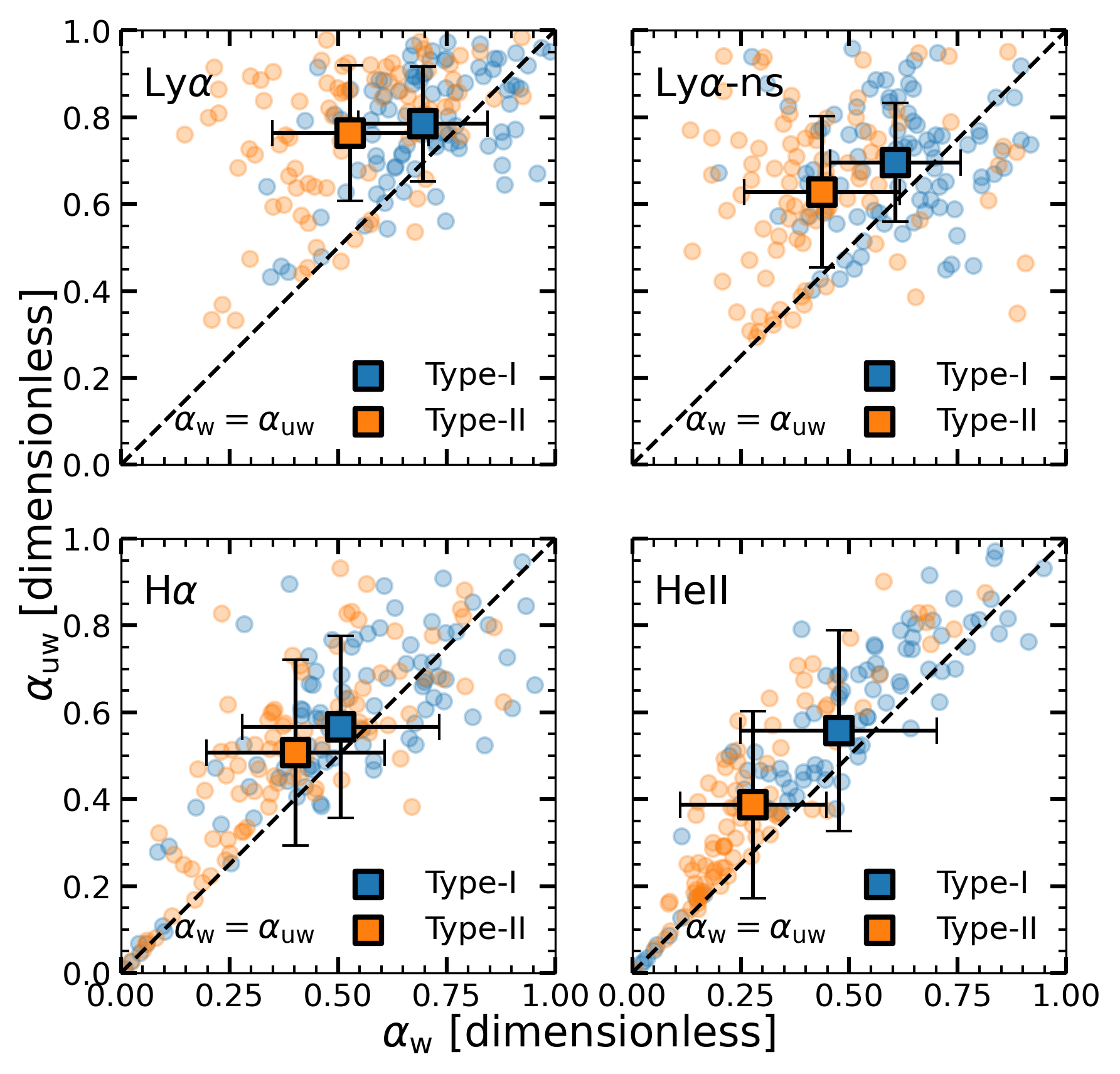}
    \caption{{\bf Upper left:} The $\alpha_{\rm w}$ versus the $\alpha_{\rm uw}$ of the Ly$\alpha$ nebulae with the processing of \textsc{rascas}. 
    The dots denote the individual simulated nebulae and the square denotes the mean value. 
    The errorbars denote the 1-$\sigma$ scatter. 
    Different colors represent different types of nebulae. 
    The dashed line represent $\alpha_{\rm w}=\alpha_{\rm uw}$. 
    {\bf Upper right, Lower left, \& Lower right:} same as the upper left panel but for the Ly$\alpha$ without the processing of \textsc{rascas}, H$\alpha$, and \heii \ nebulae.
    These panels show that the type-II nebulae are less symmetric than the type-I nebulae due to the anisotropic AGN radiation. 
    Besides, the comparison between the $\alpha_{\rm w}$ of the Ly$\alpha$ nebulae with and without the processing of \textsc{rascas} shows that the resonant scattering of the Ly$\alpha$ emission can make the nebulae more symmetric by redistributing the photons isotropically.}
    \label{nebulae_symmetry}
\end{figure}

We calculate the $\alpha_{\rm w}$ and $\alpha_{\rm uw}$ for all mock images (Fig.~\ref{nebulae_symmetry}) under the same $\Delta$ and AGN luminosity. 
The mean values of $\alpha_{\rm w}$ and $\alpha_{\rm uw}$ are shown in Tab.~\ref{sb_fitting_param_tab}. 
These results show that type-I nebulae have the higher $\alpha$ parameter, indicating that they are more symmetric than the type-II nebulae, which supports the observed result \citep{Brok2020}. 
This difference is due to the projection effect of the AGN ionization cone. 
The projection of the cone of the unobscured and obscured AGN is in circular and hourglass-like shape, respectively. 
The circular shape is more symmetric than the hourglass-like shape, making the type-I nebulae more symmetric than the type-II nebulae.

\begin{figure}
    \centering
    \includegraphics[width=\columnwidth]{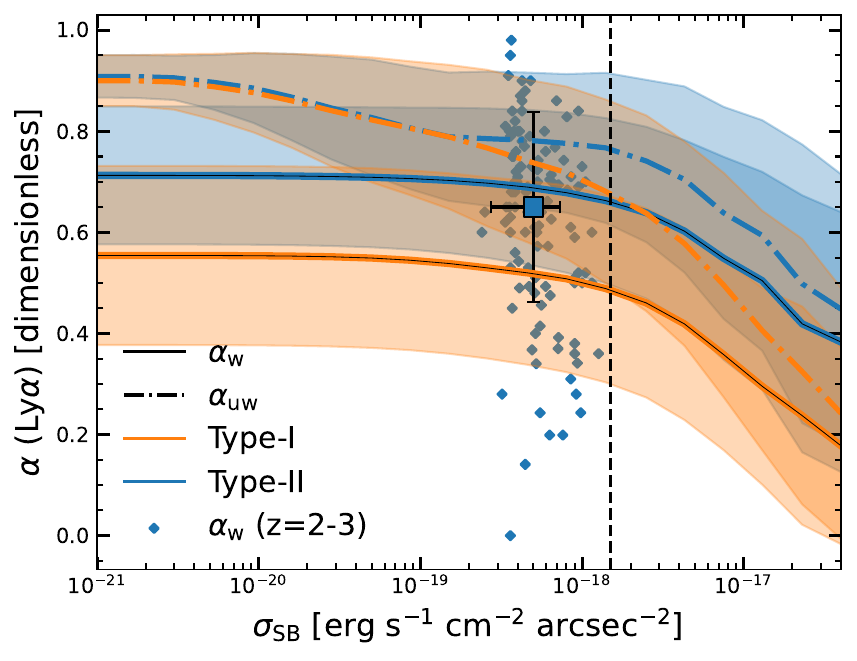}
    \caption{The relation between the SB limit ($\sigma_{\rm SB}$) and the $\alpha$ parameter of Ly$\alpha$ nebulae. 
    In a wide range of $\sigma_{\rm SB}$, the $\alpha_{\rm w}$ of the type-II nebulae keeps lower than that of the type-I nebulae while the $\alpha_{\rm uw}$ of the two types of nebulae tends to equal to each other at $\sigma_{\rm SB}\leq 3.0\times 10^{-19} {\rm erg \ s^{-1} \ cm^{-2} \ arcsec^{-2}}$, which makes $\alpha_{\rm w}$ better at characterizing the symmetry of nebulae.
    We also compare the $\alpha_{\rm w}$ (diamonds) of observed type-I Ly$\alpha$ nebulae at $z=2-3$ \citep{Cai2019,Fab2019,OSullivan2020} with the simulated results. 
    The blue square denotes the mean value of these diamonds with the errorbar denoting the 1-$\sigma$ scatter. 
    The comparison shows that our simulations could reproduce the observed $\alpha_{\rm w}$. 
    The vertical dashed line marks $\sigma_{\rm SB}=1.5\times 10^{-18} \ {\rm erg \ s^{-1} \ cm^{-2} \ arcsec^{-2}}$ where the $\alpha_{\rm }$ begins to drop faster. }
    \label{alpha_vs_noise_fig}
\end{figure}

Since the calculation of $\alpha$ is based on the selection of $\sigma_{\rm SB}$, knowing the dependence of the $\alpha$ parameter on the $\sigma_{\rm SB}$ is necessary for guiding the study of nebulae morphology in observations. 
The Ly$\alpha$ nebulae are used to reveal this dependence (Fig.~\ref{alpha_vs_noise_fig}). 
Both $\alpha_{\rm w}$ and $\alpha_{\rm uw}$ drops when the $\sigma_{\rm SB}$ increases because the faint regions where the isotropic UVB is the dominant source to power nebulae are excluded.  
Only inner regions where the anisotropic AGN radiation dominates the powering of the nebulae remain. 
In addition, the $\alpha_{\rm w}$ and $\alpha_{\rm uw}$ correlate with the $\sigma_{\rm SB}$ differently. 
The $\alpha_{\rm uw}$ of the two types of nebulae tend to be equal to each other ($\alpha_{\rm uw}\approx 0.9$) at $\sigma_{\rm SB}\leq 3.0\times 10^{-19}$ erg s$^{-1}$ cm$^{-2}$ arcsec$^{-2}$. 
The same phenomenon for the H$\alpha$ and \heii \ appears at $\sigma_{\rm SB}\leq 10^{-21}$ erg s$^{-1}$ cm$^{-2}$ arcsec$^{-2}$. 
Whereas, the $\alpha_{\rm w}$ of the type-II nebulae keeps lower than that of the type-I nebulae in a broad range ($\sigma_{\rm SB}=10^{-20} - 10^{-16}$ erg s$^{-1}$ cm$^{-2}$ arcsec$^{-2}$). 
At $\sigma_{\rm SB}\leq 1.5\times 10^{-18}$ erg s$^{-1}$ cm$^{-2}$ arcsec$^{-2}$, the $\alpha_{\rm w}$ of the two types of Ly$\alpha$ nebulae tend to be constant as $\alpha_{\rm w, I}\approx0.71$ and $\alpha_{\rm w, I}\approx0.55$. 
Since the $\alpha_{\rm w, II}$ is always lower than the $\alpha_{\rm w,I}$ in a wide range of $\sigma_{\rm SB}$ and the $\alpha_{\rm w}$ is less sensitive to the $\sigma_{\rm SB}$ than the $\alpha_{\rm uw}$, this parameter is better at quantifying the nebulae symmetry which could be used to test the AGN unified model. 
Whereas, the connection of $\alpha_{\rm w}$ to the AGN properties is required to check if $\alpha_{\rm w}$ can be used to do the test. 
This will be discussed in Sec.~\ref{property_nebulae_AGN}.

In addition, the measured $\alpha_{\rm w}$ from observed type-I Ly$\alpha$ nebulae at $z=2-3$ \citep{Cai2019,Fab2019,OSullivan2020} is also shown in Fig.~\ref{alpha_vs_noise_fig}. 
The $\alpha_{\rm w}$ of these observations take the median at $\alpha_{\rm w}=0.65\pm 0.19$ with a median $\sigma_{\rm SB}$ of $5.0\times 10^{-19} \ {\rm erg \ s^{-1} \ cm^{-2} \ arcsec^{-2}}$. 
This value is within the 1-$\sigma$ scatter of the simulated value ($\alpha_{\rm w}=0.58$) at the same $\sigma_{\rm SB}$ for the type-I nebulae. 
These comparisons indicate that our simulations could reproduce the observed symmetry of the type-I Ly$\alpha$ nebulae. 
Besides, \cite{Gillette2023} shows that the nebulae around the extremely red quasar (ERQ) at $z=2-3$ have the symmetry of $\approx 0.76$, which can be covered by our simulated type-I nebulae. 
This also favors their conclusion that the illumination pattern of the ERQ is similar to the type-I quasars. 
Moreover, Tab.~\ref{sb_fitting_param_tab} also shows that the $\alpha_{\rm w}$ parameter of Ly$\alpha$ nebulae systemically increase by $\geq 14\%$ ($\Delta \alpha_{\rm w}\approx 0.1$) after including the resonant scattering. 
This suggests that multiple random scattering events could cause the Ly$\alpha$ nebulae to be more symmetric by isotropically redistributing the Ly$\alpha$ photons.

\subsection{The surface brightness (SB) of nebulae} \label{nebulae_sb}
The SB profiles of the nebulae are valuable for studying the CGM gas properties and the AGN radiation.
In this section, we show the simulated profiles of the two types of nebulae. 
The stacked SB images are also presented. 
The cosmic dimming effect is corrected. 

\begin{figure*}
    \centering
    \includegraphics[width=.9\textwidth]{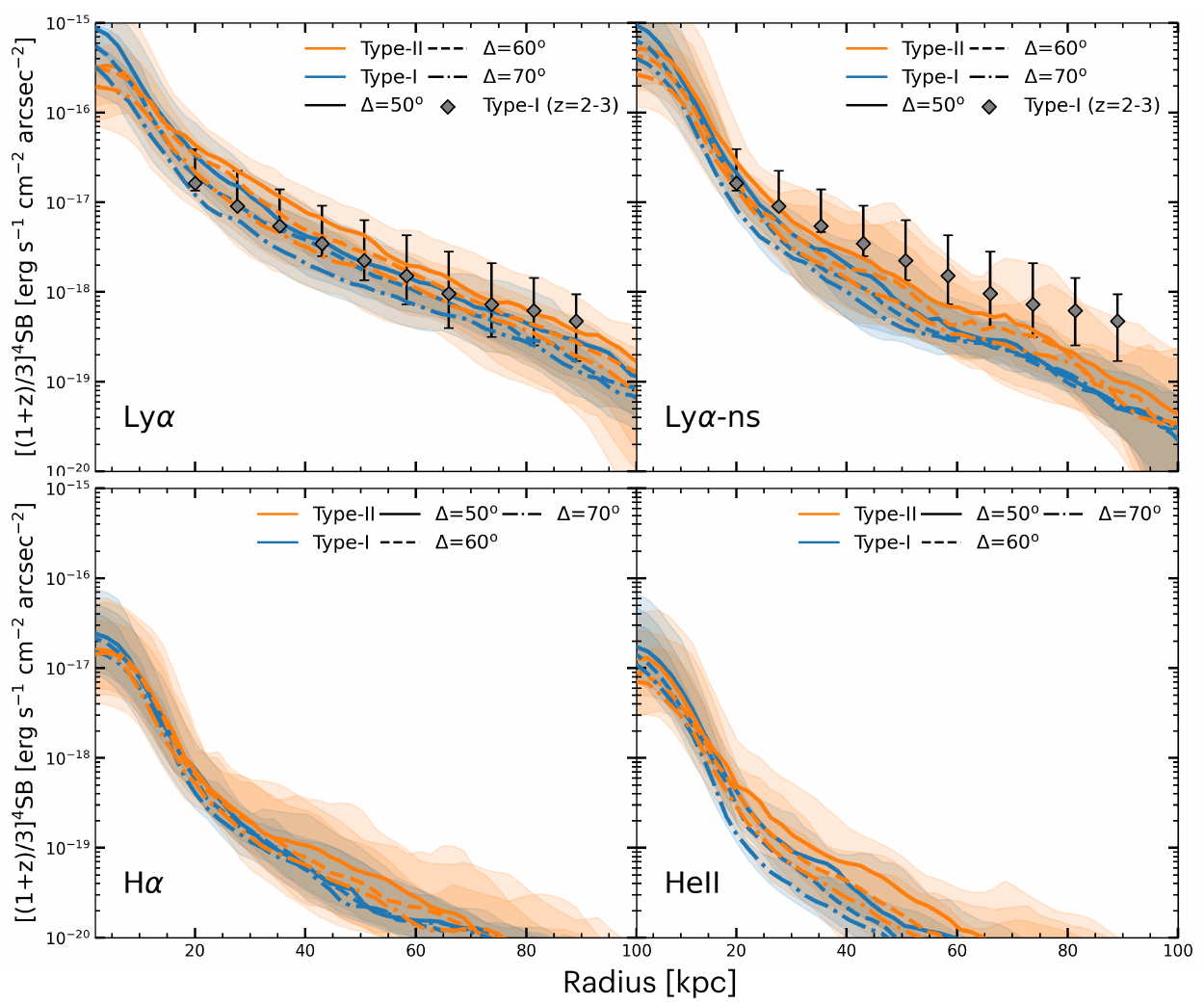}
    \caption{{\bf Upper left:} The median-stacked SB profiles of the Ly$\alpha$ nebulae with the processing of \textsc{rascas}. 
    The blue and orange colors denote the profile of type-I and type-II nebulae, respectively, with the shadows denoting the 16th and 84th percentiles. 
    The solid, dashed, dash-dotted lines denote the profile of $\Delta=50^{\rm o}$, $60^{\rm o}$, and $70^{\rm o}$, respectively. 
    The diamonds with errorbars denote the median-stacked profile of the observed type-I nebulae at $z=2-3$ \citep{Fab2019,Cai2019,OSullivan2020,Fossati2021} with the bar denoting the 16th and 84th percentiles. 
    The cosmic dimming effect is corrected to $z=2$. 
    {\bf Upper right, Lower left, \&  Lower right: } same as the upper left panel but for the Ly$\alpha$ without the processing of \textsc{rascas}, H$\alpha$, and \heii \ nebulae.  
    These plots show that the SB profiles of the type-I nebulae are flatter than those of the type-I nebulae due to different projections of the ionization cone. 
    The profile of the Ly$\alpha$ nebulae with the processing of \textsc{rascas} is flatter than that of the Ly$\alpha$ nebulae without the processing of \textsc{rascas} because the Ly$\alpha$ photons in the inner region are redistributed to the outer region by resonant scattering. 
    Moreover, the comparison between the observed Ly$\alpha$ profile with simulations favors the Ly$\alpha$ nebulae with the processing of \textsc{rascas}. 
    This indicates that the resonant scattering effect is non-negligible.}
    \label{sb_profile}
\end{figure*}

\subsubsection{The averaged radial profile} \label{sb_profile_sec}
\begin{figure*}
    \centering
    \includegraphics[width=.8\textwidth]{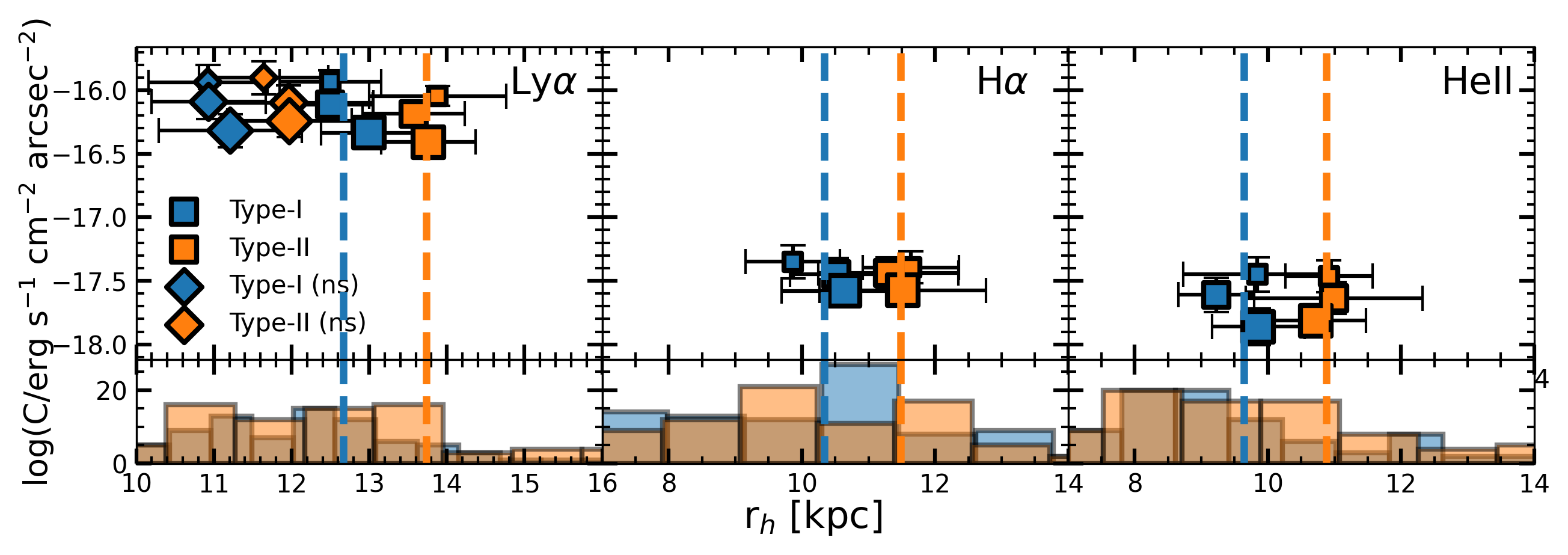}
    \caption{{\bf Left:} The mean values and histograms of the best-fit scale length ($r_{h}$) and normalization parameter ($C$) of the simulated Ly$\alpha$ nebulae. 
    The blue and orange colors denote the type-I and type-II nebulae, respectively. 
    The squares denote the mean values of the Ly$\alpha$ nebulae with the processing of \textsc{rascas} while the diamonds denote the mean values of the Ly$\alpha$ nebulae without the processing of \textsc{rascas}. 
    The sizes of these markers represent the $\Delta$ which ranges in $50^{\rm o}-70^{\rm o}$. 
    The errorbar is the 1-$\sigma$ fitting uncertainty. 
    The vertical dashed lines mark the location of the mean $r_{\rm h}$ of the squares. 
    {\bf Middle \& Right:} same as the left panel but for the H$\alpha$ and \heii \ nebulae. 
    The corresponding values in this figure are shown in Tab.~\ref{sb_fitting_param_tab}. 
    The $r_{\rm h}$ of the type-II nebulae is larger than that of the type-I nebulae, indicating that the profile of type-II nebulae is flatter.}
    \label{sb_fit_parameter}
\end{figure*}

Fig.~\ref{sb_profile} shows the radial profiles of the Ly$\alpha$, H$\alpha$, and \heii \ nebulae. 
For each $\Delta$, we apply median stacking to all 10 simulated systems and three random orientations of the ionization cone. 
All emissions exhibit the same trend: the stacked profiles of the type-II nebulae turn from lower to higher than those of type-I nebulae from the inner to the outer region at a turning radius of $\approx 10$ kpc. 
This means that the profiles of the type-II nebulae are flatter than the type-I nebulae. 
This can be quantified by the scale length ($r_{h}$) of Eq.~\ref{sb_exponential}
\begin{equation}
    {\rm SB} (r)=Ce^{-r/r_{h}}
    \label{sb_exponential}
\end{equation}
where $C$ is the normalization that measures the level of the profile and $r_{h}$ characterizes the slope of the profile. 
The best-fit parameters of the individual profile and their mean values are shown in Fig.~\ref{sb_fit_parameter} and Tab.~\ref{sb_fitting_param_tab}. 
The mean $r_{\rm h}$ of all emissions shows that the scale lengths of the type-II nebulae are higher than those of the type-I nebulae. 
We apply the one-sided Welsch t-test to the $r_{h}$ between the two types of nebulae for the three lines, which returns $p_{\rm Ly\alpha}\approx0.6\%$, $p_{\rm H\alpha}\approx2\%$, and $p_{\rm HeII}\approx0.7\%$. 
Under the threshold of $p=5\%$, the t-test confirms that the type-II nebulae have higher $r_{\rm h}$ than the type-I nebulae, i.e. type-II nebulae have flatter SB profiles than those of the type-I nebulae.
\begin{table*}
\begin{tabular}{lcccccc}
\hline
\hline
        & \multicolumn{2}{c}{Ly$\alpha$/Ly$\alpha$-ns}  & \multicolumn{2}{c}{H$\alpha$}  & \multicolumn{2}{c}{HeII}  \\ \hline
        & Type-I          &  Type-II          & Type-I          & Type-II         & Type-I           & Type-II          \\ \hline
$r_{\rm h}(\Delta=50^{\rm o})$ [kpc]   &     $12.5\pm0.7$ / $10.9\pm0.8$      &   $13.9\pm0.9$ / $11.6\pm0.8$         &  $9.9\pm0.7$         &  $11.6\pm0.7$         &  $9.8\pm1.1$           &  $10.9\pm0.7$          \\
$r_{\rm h}(\Delta=60^{\rm o})$ [kpc]      &    $12.5\pm0.6$ / $10.9\pm0.7$      &   $13.6\pm0.7$ / $12.0\pm1.1$         &  $10.5\pm0.7$          &  $11.3\pm1.1$         &  $9.2\pm0.6$           &  $11.0\pm1.3$          \\
$r_{\rm h}(\Delta=70^{\rm o})$ [kpc]      &    $13.0\pm0.6$ / $11.2\pm0.9$      &   $13.8\pm0.6$ / $12.1\pm0.8$         &  $10.6\pm0.9$          &  $11.5\pm1.2$         &  $9.9\pm0.7$           & $10.7\pm0.8$          \\
$\overline{r_{\rm h}} (\Delta)$ [kpc] &          $12.7\pm0.4$ / $11.0\pm0.5$      &   $13.7\pm0.4$ / $11.9\pm0.5$         &  $10.3\pm0.4$          &  $11.5\pm0.6$         &  $9.6\pm0.5$           &  $10.9\pm0.6$          \\ 
$\overline{\alpha_{\rm w}} (\Delta)$ & $0.70\pm0.15$ / $0.61\pm 0.15$& $0.53\pm0.18$ / $0.44\pm 0.18$ & $0.51\pm 0.23$ & $0.40\pm0.24$ & $0.48\pm0.23$ & $0.28\pm0.17$ \\
$\overline{\alpha_{\rm uw}} (\Delta)$ & $0.78\pm0.13$ / $0.70\pm 0.14$ & $0.76\pm0.16$ / $0.63\pm 0.17$ & $0.57\pm 0.21$ & $0.50\pm0.21$ & $0.56\pm0.23$ & $0.39\pm0.22$ \\
\hline
\hline
\end{tabular}
\caption{{\bf The characterizing parameters of the simulated nebulae.} 
The first three rows are the mean value of the scale length in the unit of kpc under different $\Delta$. 
The fourth row is the mean value of the three rows above. 
The errors in these four rows are the 1-$\sigma$ fitting uncertainties. 
The last two rows denote the mean value of the flux-weighted and flux-unweighted $\alpha$ parameter shown in Fig.~\ref{nebulae_symmetry}. 
The error here represents the 1-$\sigma$ scatter.}
\label{sb_fitting_param_tab}
\end{table*}

The difference between the two types of nebulae in the slope of the SB profile 
originates from a different projection of the ionization cones on the sky plane for the two types of AGN. 
When sampling the SB profile at the smallest radii, the flux of type-I nebulae is mainly contributed by gas particles within the cone (which have higher emissivities), while the flux of type-II nebulae is mainly contributed by gas out of the cone (which have lower emissivities). 
As the radius increases, fewer gas particles within the cone contribute to the flux of the type-I nebulae while more particles within the cone contribute to the flux of the type-II nebulae.
Consequently, the profiles of type-II nebulae are $\approx 3$ times lower profiles at small radii and $\approx 2$ times higher at largest radii than the type-I nebulae, making the type-II nebulae flatter. 



In addition, Fig.~\ref{sb_profile} shows that the profile of the Ly$\alpha$ without the processing of \textsc{rascas} is steeper than the Ly$\alpha$ with the processing of \textsc{rascas}. 
This is confirmed by the fitting results where the $r_{h}$ of the Ly$\alpha$ increases by $\approx 13\%$ after considering the resonant scattering effect ($r_{h}$ increases from ranging in 10.9 - 12.1 kpc to 12.5 - 13.9 kpc) because the Ly$\alpha$ photons produced at small radii propagate to large radii through multiple scatterings.  

Moreover, we also compare the observed SB profile of the type-I Ly$\alpha$ nebulae at $z=2-3$ with the simulated profiles. 
We perform the cosmic-dimming correction to the observed SB profiles \citep{Fab2019,Cai2019,OSullivan2020,Fossati2021} and apply median stacking to them. 
The simulated profile with the processing of \textsc{rascas} is within the 16th - 84th percentile of the observed profiles while the simulated profile without the processing of \textsc{rascas} is out of this percentile. 
This indicates that the profiles of Ly$\alpha$ with the processing of \textsc{rascas} are more consistent with observations than Ly$\alpha$ without the processing of \textsc{rascas}. 
This comparison suggests that the resonant scattering effect is non-negligible for Ly$\alpha$ nebulae. 
The reduced chi-square between the observed and simulated profiles are $\chi^{2}_{\rm r, 50}=0.7$, $\chi^{2}_{\rm r, 60}=0.4$, and $\chi^{2}_{\rm r, 70}=0.5$, respectively, suggesting that $\Delta=60^{\rm o}$ which corresponds to an opening angle of the ionization cone of $60^{\rm o}$ is favored.
In fact, by comparing the simulated and observed SB profile of the Ly$\alpha$ nebulae at $z=3$, \cite{Obreja2024} also favors this scenario. 
Our results and the results of \cite{Obreja2024} indicate that the unobscured AGN with the cone of $\alpha_{\rm cone}\approx 60^{\rm o}$ should compose the major population of unobscured AGNs at $z=2-3$ if the dusty torus exists. 

As for the normalization parameter $C$ which characterizes the level of the SB profile, we do not see any significant difference between the two types of nebulae (Fig.~\ref{sb_fit_parameter}). 
We will discuss the correlation between these parameters and the AGN properties in Sec.~\ref{property_nebulae_AGN}.

\subsubsection{Stacking along the axes of the torus} \label{torus_gal_stacking}
Under the AGN unified model \citep{Antonucci1993}, the AGN radio jet is commonly assumed to be perpendicular to the plane of the torus, which is supported by optical polarimetry and radio observations \citep{Vernet2001,Drouart2012,Venturi2021}.  
This indicates that the axis of the ionization cone can be found by observing the radio jet around the AGN \citep{Marques-Chaves2019,Hardcastle2020}. 
This allows us to confirm the existence of the ionization cone by stacking the image of nebulae after aligning them along the radio jet. 

\begin{figure*}
    \centering
    \includegraphics[width=\textwidth]{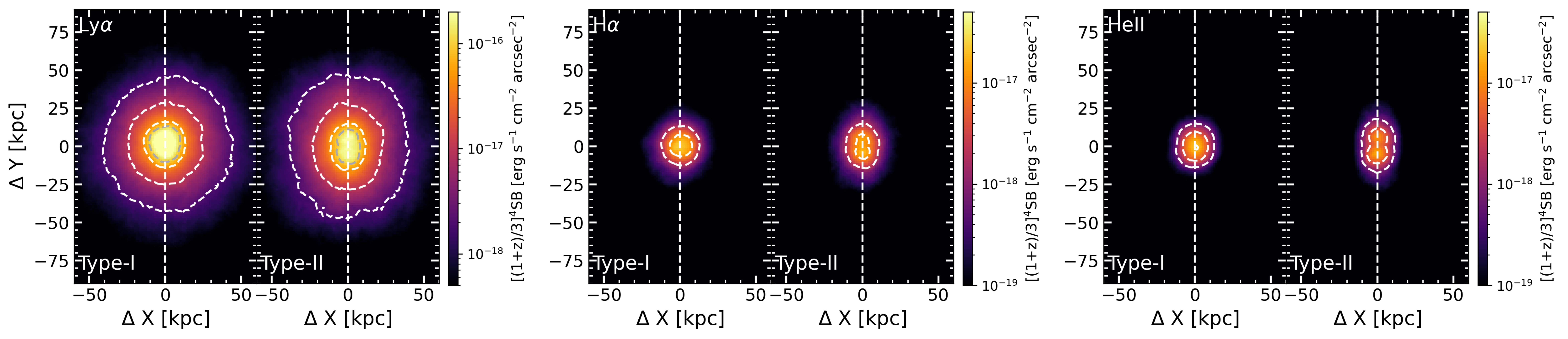}
    \caption{ {\bf Left:} the stacked images of Ly$\alpha$ nebulae after aligning 90 mock images along the torus axis for each type of nebulae. 
    The cosmic dimming effect has been corrected. 
    The vertical dashed lines represent the torus axis. 
    The dashed contours denote SB limits of [$2\sigma_{\rm SB}$, $5\sigma_{\rm SB}$, $10\sigma_{\rm SB}$, $100\sigma_{\rm SB}$] with the 1-$\sigma$ SB limit of $1.0\times 10^{-18}$ erg s$^{-1}$ cm$^{-2}$. 
    The stacked images of the type-II nebulae exhibit the obvious elongation along the torus axis while the stacked image of the type-I nebulae does not have such morphology. 
    {\bf Middle \& Right:} same as the right panel but for the H$\alpha$ and \heii \ nebulae. }
    \label{img_stack}
\end{figure*}

For this purpose, we stack 90 images for each emission and each type of nebulae by aligning them along the torus axis (Please refer to Sec.~\ref{mockobs} for the number of mock images). 
The results are shown in Fig.~\ref{img_stack}. 
The type-II nebulae of these three emissions exhibit significant elongation along the torus axis while type-I nebulae tend to be isotropic. 
For the Ly$\alpha$, the $\alpha_{\rm w}$ measured from the stacked image of the two types of nebulae are $\alpha_{\rm w, I}=0.84$ and $\alpha_{\rm w, II}=0.65$, respectively, under the SB limit of $\sigma_{\rm SB}=1.0\times 10^{-18}$ erg s$^{-1}$ cm$^{-2}$. 
For the H$\alpha$ and \heii, the $\alpha_{w}$ of the two types of nebulae are $\alpha_{\rm w, I}({\rm H}\alpha)=0.62$, $\alpha_{\rm w, II}({\rm H}\alpha)=0.11$, $\alpha_{\rm w, I}({\rm HeII})=0.51$, and $\alpha_{\rm w, II}({\rm HeII})=0.18$.
These results manifest that if enough images of type-II nebulae are stacked, the anisotropic AGN radiation, i.e. the ionization cone can be revealed, ultimately providing strong evidence for the unified model. 
We will discuss the number of type-II Ly$\alpha$ nebulae required for statistically confirming the cone under different SB limits in Sec.~\ref{nebulae_probing_AGN}.

\subsection{The Ly$\alpha$ spectral profile} \label{lya_spec}
\begin{figure*}
    \centering
    \includegraphics[width=\textwidth]{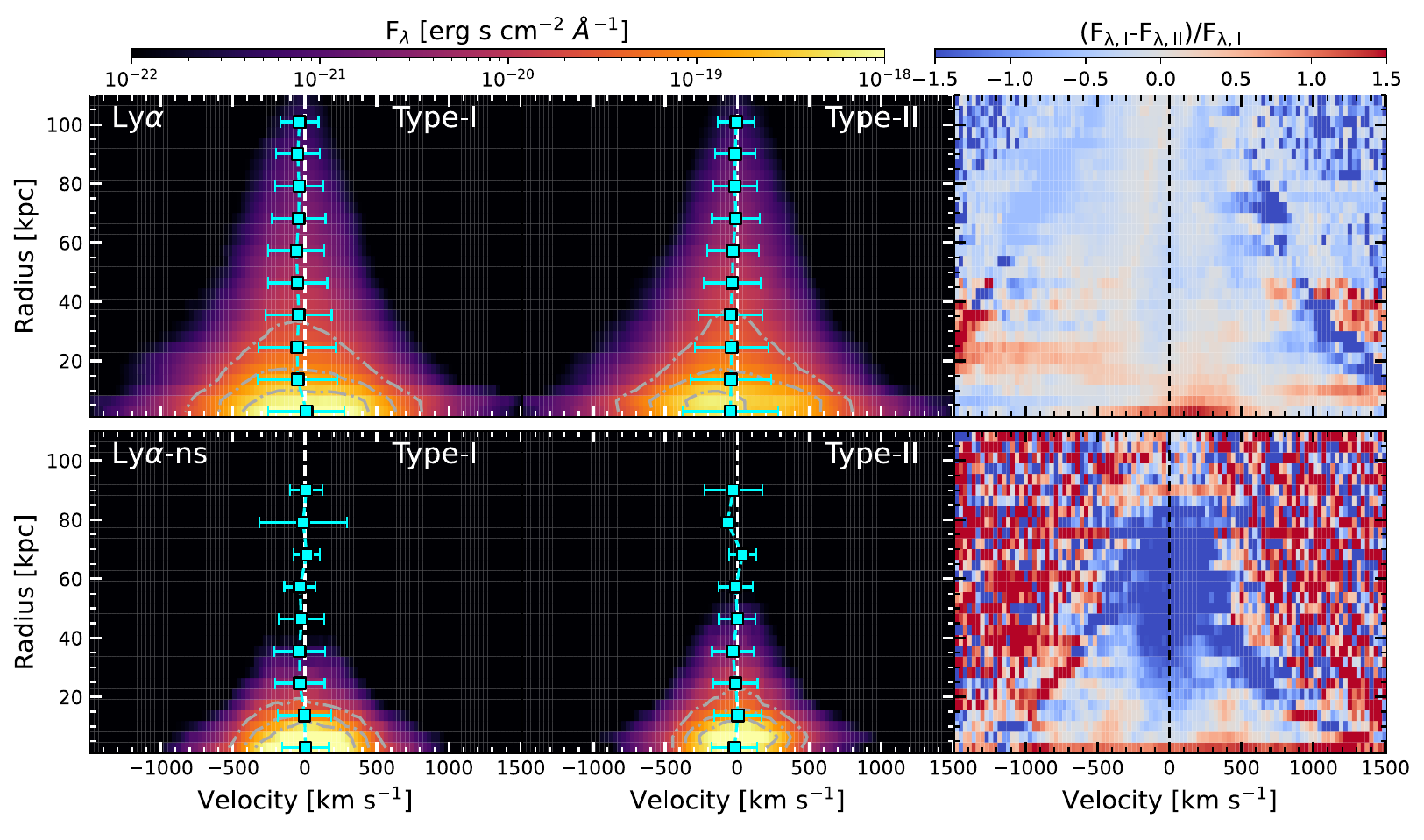}
    \caption{{\bf Upper:} The radially projected 2D spectra of the Ly$\alpha$ nebulae with the processing of \textsc{rascas}. 
    Panels from left to right are the spectra of the type-I nebulae, spectra of the type-II nebulae, and their residuals. 
    The white dashed line marks the location where $\Delta v=0$ km s$^{-1}$. 
    The dash-dotted contours represent levels of [2$\sigma$, $10\sigma$, $40\sigma$] with $\sigma=10^{-20}$ erg s$^{-1}$ cm$^{-2}$ \AA$^{-1}$. 
    The cyan squares denote the flux-weighted velocities with the errorbars denoting the flux-weighted velocity dispersions. 
    {\bf Bottom:} same as the upper panel but for the Ly$\alpha$ nebulae without the processing of \textsc{rascas}. 
    A 1D Gaussian kernel with $\rm FWHM=75 $ km s$^{-1}$ which corresponds to the resolving power of $R=4000$ is convolved to the spectra to mock line spread function (LSF) of the BM grating on Keck/KCWI. 
    The residual maps indicate that the flux density of type-I nebulae turns from higher to lower than that of the type-II nebulae from the small to the large radius, consistent with the results of Sec.~\ref{nebulae_sb}.}
    \label{lya_spec_fig}
\end{figure*}

Previous observations have shown that the spectral profiles of the emissions are necessary to analyze the gas kinematics for studying the CGM-galaxy ecosystem \citep{Chen2021,Zhang2023a,Zhang2023b}. 
In this section, we use the mock datacubes produced by the \textsc{rascas} code to show the spectra of Ly$\alpha$ nebulae with and without the processing of \textsc{rascas} to explore: (i) if the two types of nebulae exhibit any difference in spectral profile (ii) the influence of the resonant scattering effect on the Ly$\alpha$ spectra. 
180 mock datacubes of the two types of nebulae are employed in this study (Please refer to Sec.~\ref{mockobs} for the number of mock datacubes). 
For each datacube, we follow \cite{Chen2021} to produce the radially projected 2D spectra by extracting the average spectra from each radial bin. 
For each type of nebulae, we then perform the median stack to the 2D spectra and generate the final stacked 2D spectra. 
The results are shown in Fig.~\ref{lya_spec_fig}. 

\begin{figure*}
    \centering
    \includegraphics[width=\textwidth]{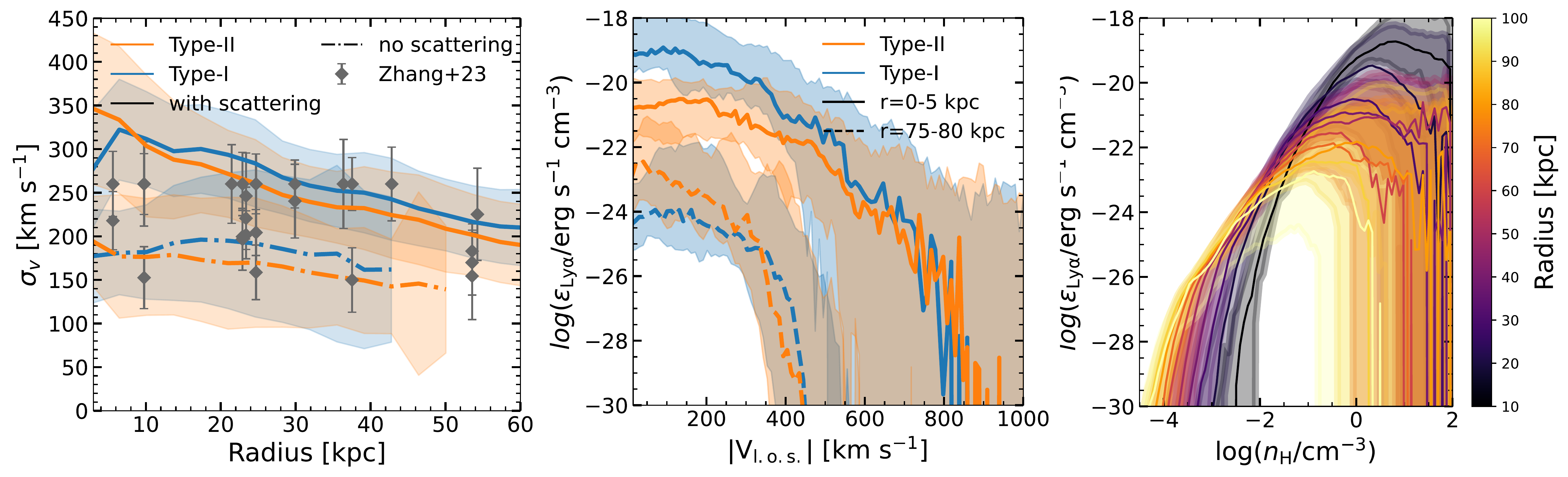}
    \caption{{\bf Left:} The radial profile of the flux-weighted velocity dispersion ($\sigma_{v}$) of the  Ly$\alpha$ nebulae with (solid line) and without (dash-dotted line) the processing of \textsc{rascas}. 
    The blue and orange lines represent the median $\sigma_{v}$ of the type-I and type-II nebulae, respectively. 
    The shadow denotes the 1-$\sigma$ scatter. 
    The diamonds represent the $\sigma_{v}$ of an observed type-II Ly$\alpha$ nebulae at $z=2.3$ \citep{Zhang2023a}. 
    This plot shows that the $\sigma_{v}$ of the type-I nebulae turns from higher to lower than that of the type-II nebulae from the small to large radius. 
    Besides, the $\sigma_{\rm v}$ of Ly$\alpha$ nebulae with the processing of \textsc{rascas} is larger than that of the Ly$\alpha$ nebulae without the processing of \textsc{rascas} by $\approx 100$ km s$^{-1}$, which is due to the resonant scattering effect. 
    {\bf Middle:} The relation between the cumulated Ly$\alpha$ emissivity ($\epsilon_{\rm Ly\alpha}$) and the absolute line-of-sight velocity. 
    The solid and dashed lines represent the emissivity in the radial bin of $0-5$ kpc and $75 - 80$ kpc, respectively with the shadows denoting the 16th and 84th percentiles. 
    The variation of emissivity of the type-I nebulae from larger to smaller than that of the type-II nebulae is the reason for the variation of the radial profile of $\sigma_{v}$. 
    {\bf Right:} The relation between the cumulated Ly$\alpha$ emissivity and $n_{\rm H}$ at different radii (color coded). 
    The $n_{\rm H}$ that contributes most to the Ly$\alpha$ emissivity extends to the lower value as the radius increases.}
    \label{lya_specdisp_fig}
\end{figure*}

The spectra of the type-II Ly$\alpha$ nebulae are slightly different from those of the type-I nebulae. 
The residual maps exhibit that the flux density of the type-I nebulae turns from higher to lower than that of the type-II nebulae from small to large radii, consistent with the SB profiles (Fig.~\ref{sb_profile}). 
At $r\leq 10$ kpc, the spectral profile of the type-I nebulae centers at $\Delta v\approx 0$ km s$^{-1}$ while the profile of the type-II nebulae centers at $\Delta v \approx -100$ km s$^{-1}$. 
The blueshift is due to the scattering because the type-II nebulae without the processing of \textsc{rascas} do not show this phenomenon. 
The radial profile of the line width which is characterized by the flux-weighted velocity dispersion ($\sigma_{v}$) is shown in Fig.~\ref{lya_specdisp_fig}. 
For the Ly$\alpha$ nebulae with and without the processing of \textsc{rascas}, the $\sigma_{v}$ of the type-I nebulae turns from lower by $\approx 20 - 70 \ {\rm km \ s^{-1}}$ to higher by $\approx 20 \ {\rm km \ s^{-1}}$ than the type-II nebulae out to the largest radius. 

This result is a natural consequence of the emissivity varying with radius. 
Fig.~\ref{lya_specdisp_fig} shows the emissivity of Ly$\alpha$ without the processing of \textsc{rascas} cumulated in different line-of-sight (l.o.s.) velocity bins where the velocity denotes the absolute value. 
For the radial bin of $0-5$ kpc, i.e. small radius, the emissivity of the type-I Ly$\alpha$ nebulae turns from higher to consistent with the emissivity of the type-II nebulae at the velocity of $\approx 700$ km s$^{-1}$. 
This causes the gas with the velocity of $|v_{\rm l.o.s.}| \leq 700$ km s$^{-1}$ around the unobscured AGN to have a higher weight than the gas around the obscured AGN. 
The type-I nebulae, thus, have larger $\sigma_{v}$ than the type-II nebulae at the inner region. 
Since the situation is reversed at the radial bin of $75-80$ kpc, i.e. large radius, the $\sigma_{v}$ of the type-I nebulae becomes smaller than that of the type-II nebulae. 
We note that detecting this small offset in $\sigma_{v}$ ($\approx 20$ km s$^{-1}$) between the two types of Ly$\alpha$ nebulae at the large radius is beyond the ability of current instruments.

In addition, Fig.~\ref{lya_specdisp_fig} also shows that the Ly$\alpha$ with the processing of \textsc{rascas} have higher $\sigma_{v}$ than the Ly$\alpha$ without the processing of \textsc{rascas} by $\approx 100 \ {\rm km \ s^{-1}}$. 
This is due to the resonant scattering which could redistribute photons from the line center to the wing in the velocity space. 
Such phenomenon was also shown by \cite{Costa2022}. 
The radial profile of velocity dispersion of an ELAN \citep{Zhang2023a} is also shown in Fig.~\ref{lya_specdisp_fig}. 
This observed velocity dispersion ranging in $150 -250$ km s$^{-1}$ is between the profiles of the Ly$\alpha$ with and without the processing of \textsc{rascas}. 
This could indicate that the resonant scattering effect does not fully dominate the spectral profile of the Ly$\alpha$ in this ELAN.

\subsection{The line ratio of \heii/Ly$\alpha$} \label{HeII_Lya_ratio} 
\cite{Lau2022} suggests that the type-I nebulae should have a lower \heii/Ly$\alpha$ ratio than the type-II nebulae under the AGN unified model when only recombination radiation is included, making this line ratio a potential indicator for distinguishing the two types of nebulae. 
This originates from the fact that the \heii/Ly$\alpha$ drops with radius because the \heii \ flux declines with distance from the AGN more steeply than the Ly$\alpha$ flux \citep{Cantalupo2019}. 
At the same projected radius, ionized gas projected onto the sky plane under the type-I case spans a larger physical distance in the ionization cone than the gas under the type-II case. 
This will cause the type-I nebulae to have smaller \heii/Ly$\alpha$ than the type-II nebulae.





\begin{figure*}
    \centering
    \includegraphics[width=\textwidth]{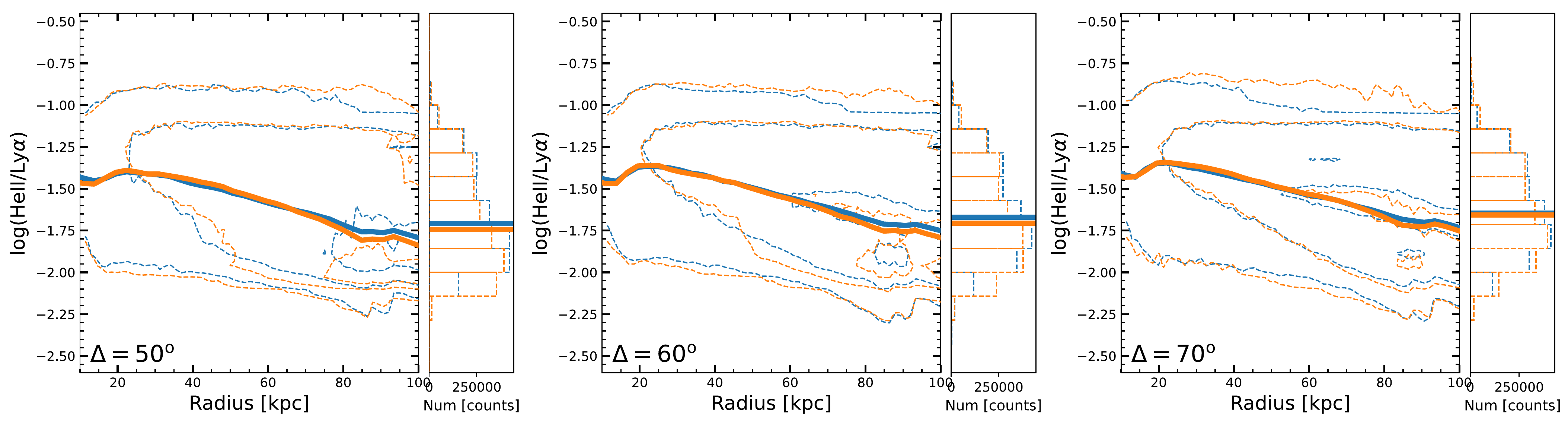}
    \caption{The radial profiles of \heii/Ly$\alpha$ under different opening angle of dust torus ($\Delta$). 
    These profiles are generated by collecting the pixels from all mock ratio maps of the 10 massive systems (90 mock maps in total).
    The blue and orange solid lines denote the median profile of the type-I and type-II nebulae, respectively. 
    The dashed contours denote the number of pixels of 100, 1000, and 3000.  
    Our simulations show that the two types of nebulae should have a consistent profile of \heii/Ly$\alpha$ in $10-100$ kpc when they share the same intrinsic luminosity under the AGN unified model. }
    \label{heii_lya_profile}
\end{figure*}

To check whether or not the type-I nebulae have lower \heii/Ly$\alpha$ than the type-II nebulae, we collect the pixels from the mock ratio maps of the 10 massive systems to generate the radial profiles of \heii/Ly$\alpha$ and show them in Fig.~\ref{heii_lya_profile}. 
Our simulations suggest that the two types of nebulae have roughly the same \heii/Ly$\alpha$ which drops from $\approx0.03$ to $\approx0.02$ in $10-100$ kpc. 
The inconsistency between our results and the discussion in \cite{Lau2022} is due to the gas outside the cone, which also plays a significant role in regulating the \heii/Ly$\alpha$. 
From our simulations, the volume outside the cone contains up to $95\%$ of the total gas particles that can contribute up to $96\%$ of the total flux for the Ly$\alpha$. 
Since the effect introduced by \cite{Cantalupo2019,Lau2022} also happens to the gas in this volume, the \heii/Ly$\alpha$ of type-I (type-II) nebulae will increase (decrease) when this volume is included, resulting in the two types of nebulae having consistent \heii/Ly$\alpha$. 
Whereas, when focusing solely on regions illuminated by the AGN—namely, those within the ionization cone—the \heii/Ly$\alpha$ ratio in type-II nebulae is $\approx 0.25-0.75$ dex higher than that in type-I nebulae at scales of $10-100$ kpc. 
This is a qualitatively consistent result with the findings of \cite{Lau2022} due to the effect introduced by \cite{Cantalupo2019}.

It is noteworthy that the comparison between our simulations and the literature results is
not straightforward for several reasons. 
First, our simulations do not include gas on Mpc scales, which could influence the \heii/Ly$\alpha$ at projected distances of tens of kpc \citep{Cantalupo2019}. 
Second, previous observations suggest that both the Ly$\alpha$ and \heii \ are correlated with the AGN feedback \citep{Reines2013,Moran2014,Liu2024}, which is not implemented in our simulations. 
Third, we assume that the two types of AGNs share the same intrinsic luminosity (Sec.~\ref{agn_sed}), whereas it remains uncertain whether the observed populations of these two AGN types truly share similar intrinsic luminosities. 
Therefore, simulations that incorporate AGN feedback and span larger volumes are necessary to investigate how the \heii/Ly$\alpha$ varies with AGN luminosity, and to draw more robust conclusions about its potential correlation with the AGN type under the unified model. 

As for observations, \cite{Lau2022} demonstrates that the type-II nebulae have larger \heii/Ly$\alpha$ ($\approx 0.1$) than the type-I nebulae (\heii/Ly$\alpha\approx 0.01$)  \citep{Cai2017,Marino2019,Brok2020,Guo2020,Sanderson2021}. 
Whereas, some observations suggest that the \heii/Ly$\alpha$ of type-I nebulae could be roughly equal or even higher than $0.1$ in some parts of the structure \citep{Borisova2016,Fab2018,Cantalupo2019,Fossati2021}. 
Moreover, only a few cases detect the Ly$\alpha$ and \heii \ nebulae simultaneously \citep{Cai2017,Brok2020,Sanderson2021}, causing the comparison in \cite{Lau2022} to be less statistical. 
Thus, whether or not the type-II nebulae have higher \heii/Ly$\alpha$ than the type-I nebulae is debated. 
More observations targeting the \heii \ nebulae are required for reaching a solid result. 

\section{Discussion} \label{sim_discussion}

In this section, we explore how the results shown in Sec.~\ref{sim_results} correlate with the AGN properties (Sec.~\ref{property_nebulae_AGN}) and how to reveal the AGN ionization cone to probe the unified model at $z=2-3$ (Sec.~\ref{nebulae_probing_AGN}). 
Since the AGN feedback is not involved in the selected halos, we will discuss the influence of the AGN feedback in Sec.~\ref{open_question}. 

\subsection{Relation between the nebulae and AGN properties} \label{property_nebulae_AGN}

Various observations have revealed the connection between CGM nebulae and their host AGN \citep{Fab2019,Mackenzie2021,Daddi2022,Gonz2023}. 
Recent simulations exhibit that the tight correlation between the nebulae and their host AGN properties can be used to constrain the AGN engine \citep{Obreja2024}. 
Nevertheless, these works focused on the type-I nebulae while these correlations for the type-II nebulae have not been studied yet. 
This section explores how the AGN intrinsic luminosity ($L_{\rm AGN}$) and the torus half-opening angle ($\Delta$) connect to the morphology and SB profile of nebulae. 
Only the A$_{1}$ massive system (Tab.~\ref{halo_info}) at $z=3.0$ is used in this analysis to save time.



\subsubsection{The correlation with the half-opening angle of the torus} \label{corr_with_oa}
\begin{figure*}
    \centering
    \includegraphics[width=\textwidth]{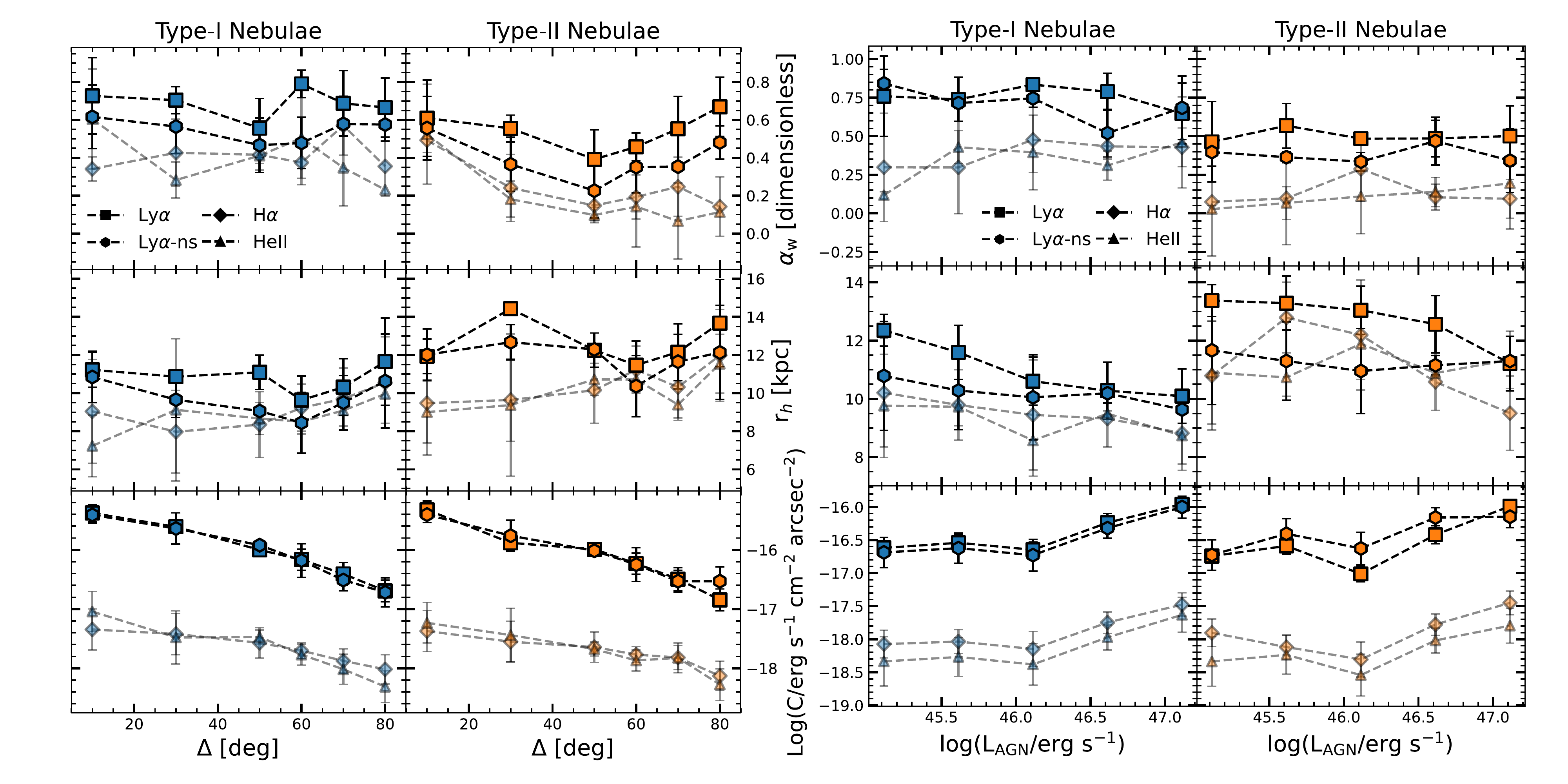}
    \caption{{\bf Left:} correlations between the nebulae properties ($\alpha_{\rm w}$, $r_{h}$, and $C$) and the half-opening angle of the torus ($\Delta$) of the two types of nebulae. 
    The square, hexagon, diamond, and triangle represent the Ly$\alpha$ with the processing of \textsc{rascas}, Ly$\alpha$ without the processing of \textsc{rascas}, H$\alpha$, and \heii, respectively. 
    The errorbar denotes the 1-$\sigma$ scatter. 
    No significant correlation is shown between $\alpha_{\rm w}$ and $\Delta$ for the type-I nebulae. 
    For the type-II nebulae, this correlation is non-linear if the SB threshold is low enough. 
    No nebula exhibits correlation between $r_{h}$ and $\Delta$ while all nebulae exhibit significant anti-correlation between $C$ and $\Delta$. 
    {\bf Right:} same as the left panel but for the correlations between nebulae properties and the intrinsic AGN luminosity ($L_{\rm AGN}$). 
    No nebulae exhibit a significant correlation between $\alpha_{\rm w}$ and $L_{\rm AGN}$. 
    The Ly$\alpha$ nebulae exhibit a significant anti-correlation between $r_{h}$ and $L_{\rm AGN}$ while other nebulae do not. 
    The correlations between the $C$ and $L_{\rm AGN}$ are significant for all nebulae. 
    These plots indicate that the correlations between the nebulae and AGN are complicated, making testing the unified model with only nebulae observations impossible.}
    \label{nebulae_vs_agn}
\end{figure*}

We let the $\Delta$ vary in the range [$10^{\rm o}$, $30^{\rm o}$, $50^{\rm o}$, $60^{\rm o}$, $70^{\rm o}$, $80^{\rm o}$]. 
This range corresponds to the opening angle of the ionization cone of $20^{\rm o} - 160^{\rm o}$. 
We randomly select three orientations for each $\Delta$ to put the ionization cone. 
The above presetting leads to $6\times 3\times 2=36$ mock images for each emission line. 
The A$_{1}$ system has a SMBH mass of $M_{\rm BH}=10^{8.75} \ M_{\odot}$ which corresponds to $L_{\rm AGN}\approx7.1\times 10^{45} \ {\rm erg \ s^{-1}}$. 
The correlations between the $\Delta$ and the nebulae properties ($\alpha_{\rm w}$, $r_{h}$, and normalization, $C$) are shown in Fig.~\ref{nebulae_vs_agn}. 
The Spearman rank correlation coefficient ($R_{s}$) is applied to quantify the significance of these correlations (Tab.~\ref{PCC_value}). 
The $R_{s}$ ranges from -1 to 1 with $R_{s}$=1 and $R_{s}$=-1 denoting the perfect positive and negative linear correlation, respectively. 
The p-value represents the chance that the correlation between the two variables is true. 
We set $p\geq 95$\% as the threshold. 


\begin{table*}
\begin{tabular}{lcccccc}
\hline
\hline
        & \multicolumn{2}{c}{Ly$\alpha$}  & \multicolumn{2}{c}{H$\alpha$}  & \multicolumn{2}{c}{HeII}  \\ \hline
        & Type-I          & Type-II          & Type-I          & Type-II         & Type-I           & Type-II          \\ \hline
$R_{s}$ ($\Delta$, $\alpha_{\rm w}$) / $p$-value [\%] & -0.08 / 24 & 0.01 / 4 & -0.19 / 56 & -0.63 / $\geq 99$ & -0.32 / 76 & -0.67 / $\geq 99$ \\
$R_{s}$ ($\Delta$, ${\rm log} (C)$) / $p$-value [\%] & -0.95 / $\geq 99$ & -0.90 / $\geq 99$ & -0.71 / $\geq 99$ & -0.78 / $\geq 99$ & -0.83 / $\geq 99$ & -0.74 / $\geq 99$ \\
$R_{s}$ ($\Delta$, $r_{h}$) / $p$-value [\%] & -0.08 / 24 & 0.02 / 1 & 0.31 / 78 & 0.35 / 85 & 0.38 / 88 & 0.30 / 60 \\
\hline
$R_{s}$ ($L_{\rm AGN}$, $\alpha_{\rm w}$) / p-value [\%] & -0.16 / 44 & $\approx 0.0$ / 2 & 0.27 / 67 & 0.06 / 16 & 0.38 / 79 & 0.37 / 81 \\
$R_{s}$ ($L_{\rm AGN}$, ${\rm log} (C)$) / p-value [\%] &0.95 / $\geq 99$ & 0.89 / $\geq 99$ & 0.85 / $\geq 99$ & 0.84 / $\geq 99$ & 0.78 \ $\geq 99$ & 0.84 \ $\geq 99$ \\
$R_{s}$ ($L_{\rm AGN}$, $r_{h}$) / p-value [\%] &-0.74 / $\geq 99$ & -0.53 / $\geq 99$ & -0.15 / 80 & -0.18 / 87 & 0.10 / 75 & 0.06 / 17 \\
\hline
\hline
\end{tabular}
\caption{{\bf {\rm $R_{s}$} and the corresponding p-value of correlations between AGN and nebulae.} 
The first three rows show the $R_{s}$ of correlations between the half-opening angle of the torus ($\Delta$) and nebulae properties and the corresponding p-value. 
The last three rows show the $R_{s}$ of correlations between the $L_{\rm AGN}$ and nebulae properties and the corresponding p-value.}
\label{PCC_value}
\end{table*}
{\bf (i) For $\Delta$ versus $\alpha_{\rm w}$:} all type-I nebulae (Ly$\alpha$, H$\alpha$, and \heii) have $\rm |R_{s}|\leq 0.32$ with the p-value of $p\leq80$\%, indicating that the correlation between the $\alpha_{\rm w}$ and $\Delta$ is not significant. 
For the type-II nebulae, the $\alpha_{\rm w}$ of the Ly$\alpha$ does not show a significant linear correlation with $\Delta$ either, while the H$\alpha$ and \heii \ exhibit significant linear anti-correlation with $\rm R_{s}\leq-0.63$ and $p\geq$99\%. 

These correlations and non-correlations originate from the variation of the cone projection with $\Delta$. 
Since the sightline is within the ionization cone for the type-I nebulae, the projection of the cone on the sky plane remains approximately circular for different $\Delta$. 
This causes that $\alpha_{\rm w}$ hardly changes with $\Delta$. 
For the type-II nebulae, the correlation is non-linear. 
When the $\Delta$ increases from $0^{\rm o}$ to $50^{\rm o}$, the volume covered by the ionization cone shrinks which makes the ionizing radiation field more anisotropic. 
This reduces the $\alpha_{\rm w}$. 
When the $\Delta$ increases from $50^{\rm o}$ to $\approx 90^{\rm o}$ (no ionization cone), the volume converts from being fractionally covered by the AGN ionization cone to free from the cone. 
This makes the ionizing radiation field return to isotropic (only the radiation from the host galaxy and UVB remains). 
The $\alpha_{\rm w}$, thus, increases. 
The continued decreasing $\alpha_{\rm w}$ of the H$\alpha$ and \heii \ at $\Delta \geq 50^{\rm o}$ is because only regions within the ionization cone are bright enough beyond the threshold of $10^{-19}$ erg s$^{-1}$ cm$^{-2}$ arcsec$^{-2}$ to be recognized as part of the nebulae. 
The $\alpha_{\rm w}$ of the H$\alpha$ and \heii \ will also rise at $\Delta \geq 50^{\rm o}$ if we loose the SB limit to $10^{-23}$ erg s$^{-1}$ cm$^{-2}$ arcsec$^{-2}$.










{\bf (ii) For $\Delta$ versus $r_{h}$:} Tab.~\ref{PCC_value} shows that the $r_{h}$ of the H$\alpha$ and \heii \ have weak correlation with $\Delta$ with the $\rm R_{s} \geq0.30$ and $p\geq 60$\%.  
This result means that the SB profiles of the H$\alpha$ and \heii \ tend to become flatter as the $\Delta$ increases (the opening angle of the ionization cone decreases). 
\cite{Obreja2024} found the same phenomenon for the \heii \ by simulating the type-I nebulae at $z=3$. 
As for the Ly$\alpha$, the two types of nebulae have the $\rm |R_{s}|\leq 0.1$ and the $p\leq 24$\%, suggesting no significant correlation between $r_{h}$ and $\Delta$.

{\bf (iii) For $\Delta$ versus ${\rm log} (C)$:} the two types of nebulae of all emissions (Ly$\alpha$, H$\alpha$, and \heii) exhibit a strong anti-correlation between ${\rm log}(C)$ and $\Delta$ with $\rm R_{s}\leq -0.7$ and $p\geq 99$\%. 
This is a natural consequence of the volume of the region illuminated by the AGN shrinking as $\Delta$ increases (the opening angle of the cone decreases). 
We find that the nebulae could become $\sim 16$ times brighter when $\Delta$ decreases from $80^{\rm o}$ to $10^{\rm o}$. 
A similar effect was seen in \cite{Obreja2024} for the type-I nebulae at $z=3$.

\subsubsection{The correlation with the AGN intrinsic luminosity} \label{corr_with_Mbh}
In this section, we fix $\Delta$ to $60^{\rm o}$ to explore the relation between the AGN intrinsic luminosity and the nebulae properties. 
We let the $M_{\rm BH}$ vary in $10^{8}-10^{10} \ M_{\odot}$ in a step of 0.5 dex.  
These values of $M_{\rm BH}$ corresponds to AGN intrinsic luminosities of $L_{\rm AGN}=10^{45.1} - 10^{47.1}$ erg s$^{-1}$ under the fixed Eddington ratio of $\lambda=0.1$. 
Three random orientations are selected for each mass bin to put the ionization cone. 
These settings provide us with 30 mock images for each emission line. 
We also employ the $R_{s}$ to quantify the correlations. 
The results are shown in Fig.~\ref{nebulae_vs_agn} and Tab.~\ref{PCC_value}.

{\bf (i) For $L_{\rm AGN}$ versus $\alpha_{\rm w}$:} No significant correlation is found for the two types of nebulae of all emissions (Ly$\alpha$, H$\alpha$, and \heii). 
For $L_{\rm AGN}$ ranges in $10^{45.1}  - 10^{47.1} \ {\rm erg \ s^{-1}}$, the $\alpha_{\rm w}$ of the type-I Ly$\alpha$ nebulae ranges in 0.58 - 0.76 and the $\alpha_{\rm w}$ of the type-II nebulae ranges in 0.46 - 0.56.  

{\bf (ii) For $L_{\rm AGN}$ versus $r_{h}$:} The Ly$\alpha$ nebulae show the significant anti-correlation with $\rm R_{s}\leq -0.53$ and the p-value of $\geq 99$\%. 
The H$\alpha$ and \heii \ have the $|R_{s}|\leq0.18$ and the p-value of $\leq 87\%$, indicating the that correlation between the $r_{h}$ and the AGN intrinsic luminosity is not significant.

{\bf (iii) For $L_{\rm AGN}$ versus ${\rm log}(C)$:} Fig.~\ref{nebulae_vs_agn} and Tab.~\ref{PCC_value} confirms that the ${\rm log}(C)$ of all emissions positively correlate with the AGN intrinsic luminosity with the $|R_{s}|\geq 0.78$ and the p-value of $\geq 99\%$. 
This correlation is consistent with \cite{Obreja2024} which simulates the type-I nebulae at $z=3$. 
%


This whole section (Sec.~\ref{property_nebulae_AGN}) indicates that the morphology and SB profiles of nebulae are tightly connected to the half-opening angle and intrinsic luminosity of the AGN. 
On the one hand, these relations allow the nebula properties to be used to constrain the AGN properties. 
For example, the $r_{h}$ of the Ly$\alpha$ nebulae can be used to constrain the AGN intrinsic luminosity because they follow the monotonic anti-correlation. 
On the other hand, testing the AGN unified model by only observing nebulae is impossible due to these correlations. 
Because testing the AGN unified model by comparing the nebulae properties requires the $\Delta$ and $L_{\rm AGN}$ to be very well controlled. 
For example, if the $\Delta$ is not controlled, the type-II Ly$\alpha$ nebulae could have larger $\alpha_{\rm w}$ with $\Delta=10^{\rm o}$ ($\alpha_{\rm w}=0.61\pm0.07$) than the $\alpha_{\rm w}$ type-I Ly$\alpha$ nebulae with $\Delta=50^{\rm o}$ ($\alpha_{\rm w}=0.56\pm0.15$). 
Whereas, the $\Delta$ cannot be measured from observations if the AGN unified model is not probed. 
Besides, testing the unified model through the $r_{h}$ of the nebulae requires the precise measurement on both the $r_{h}$ and $L_{\rm AGN}$ which is beyond the ability of current instruments.

\subsection{Revealing the AGN ionization cone to probe the unified model at $z=2-3$} \label{nebulae_probing_AGN}
Since only nebulae observations are not enough to test the AGN unified model, we discuss if the unified model can be probed by revealing the ionization cone. 
Sec.~\ref{torus_gal_stacking} shows that the cone can be detected if enough images of the type-II Ly$\alpha$ nebulae are stacked by aligning them along the radio jets. 
This is based on the assumption that the radio jets are coaxial with the torus which has been probed by previous observations \citep{Vernet2001,Drouart2012}. 
Since recently the jets have been routinely discovered around radio-quiet AGN \citep{Girdhar2022,Girdhar2024,Singha2023} in addition to the radio-loud AGN \citep{Nesvadba2006,Nesvadba2008,Mullaney2013}, testing the AGN unified model by the joint observations of Ly$\alpha$ nebulae and radio jets around AGN is possible. 
We note that the stacked type-II Ly$\alpha$ nebulae could also appear elongated if the Ly$\alpha$ emissivity is enhanced by jets along their axis under an isotropic ionizing radiation field. 
However, this scenario can be ruled out if the stacked type-I Ly$\alpha$ nebulae do not exhibit equal asymmetry to the type-II Ly$\alpha$ nebulae. 
This indicates that the elongated stacked type-II nebulae and non-elongated type-I nebulae pinpoint the AGN ionization cone. 
This section will explore the required sample size to statistically confirm the cone under different SB limits. 
This will help us understand if current instruments are able to reveal the cone.



We let the SB limit vary in $10^{-21} - 10^{-17}$ erg s$^{-1}$ cm$^{-2}$ arcsec$^{-2}$ with a step of $\approx 0.08$ dex and the sample size vary in $1 - 100$ in a step of 1. 
For a certain combination of the sample size of $n_{i}$ and SB limit $\sigma_{\rm SB,i}$, we sample mock images 100 times with replacement. 
We stack images by aligning the nebulae along the torus axis for each iteration. 
The $\alpha_{\rm w}$ is calculated to quantify the elongation of the two types of nebulae. 
The fraction of $\alpha_{\rm w, II}$ below $\alpha_{\rm w,I}-3\sigma_{\rm \alpha, I}$ represents the chance that the stacked type-II nebulae show significant elongation than the stacked type-I nebulae. 
The $\alpha_{\rm w, I}$, $\alpha_{\rm w, II}$, and $\sigma_{\rm \alpha, I}$ denote the $\alpha_{\rm w}$ measured from the stacked images of the two types of nebulae and the $1\sigma$ scatter of $\alpha_{\rm w, I}$, respectively. 
We loop this process for every combination of the $n_{i}$ and $\sigma_{\rm SB,i}$ to generate the 2D map of the fraction. 
We take $f=95\%$ as the threshold that the elongation of the type-II nebulae is statistically significant.



\begin{figure}
    \centering
    \includegraphics[width=\columnwidth]{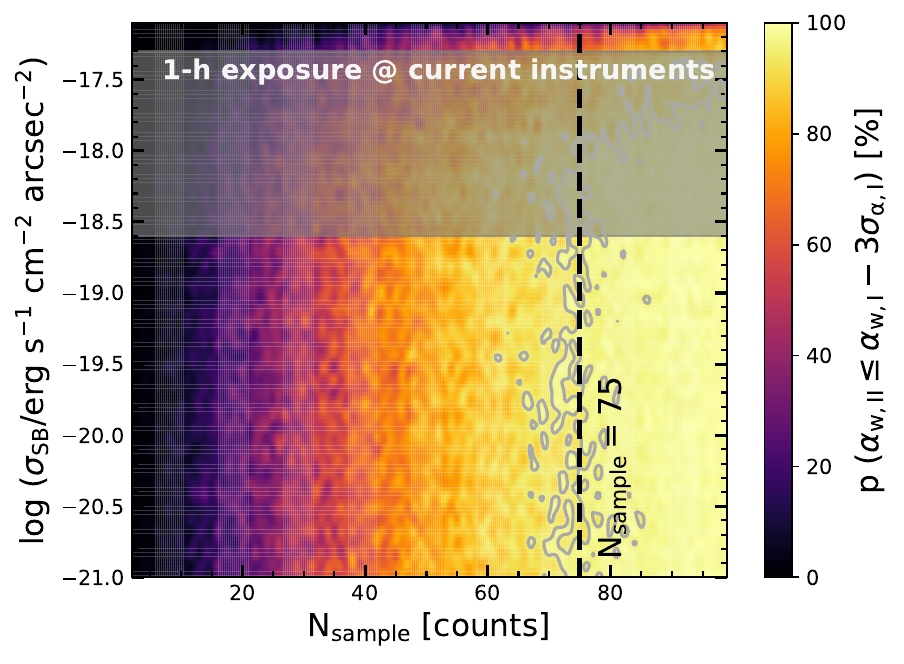}
    \caption{The probability map showing the chance that the stacked type-II Ly$\alpha$ nebulae have the $\alpha_{\rm w}$ below the 3-$\sigma$ scatter of the $\alpha_{\rm w}$ of the stacked type-I Ly$\alpha$ nebulae. 
    The $x$-axis represents the sample size and the $y$-axis represents the 1-$\sigma$ SB limit. 
    The colorbar represents the fraction. 
    The shadow represents the SB limit yield by the 1-h exposure with current facilities \citep{Fab2019,Cai2019,OSullivan2020,Li2024}. 
    The gray contour represents the fraction of 95\%. 
    Our results indicate that $\geq 75$ type-II nebulae (vertical dashed line) with the detection of radio jets are required to reveal the ionization cone under a confidence level of 95\% by using current instruments.}
    \label{probability_stack}
\end{figure}

The prediction is shown in Fig.~\ref{probability_stack}. 
At the SB limit of $\leq 10^{-17.8}$ erg s$^{-1}$ cm$^{-2}$ arcsec$^{-2}$, $\geq 75$ sources with the detection of radio jets for each type of nebulae are required for statistically confirming that the type-II nebulae are more elongated than the type-I nebulae, i.e. pinpointing the AGN ionization cone. 
This sample size will rapidly increase when $\sigma_{\rm SB}\geq 10^{-17.8}$ erg s$^{-1}$ cm$^{-2}$ arcsec$^{-2}$. 
The threshold of this jump is associated with the brightness of the nebulae. 
In Sec.~\ref{nebulae_asymmetry}, we have already shown that the $\alpha_{\rm w}$ of the two types of nebulae will decrease rapidly at $\sigma_{\rm SB}\geq 10^{-17.8}$ erg s$^{-1}$ cm$^{-2}$ arcsec$^{-2}$ (Fig.~\ref{alpha_vs_noise_fig}). 

Previous works show that current state-of-art IFS and NB imaging including the MUSE/VLT, KCWI/Keck, Palomar Cosmic Web Imager on the Hale Telescope, and the Hyper Suprime-Cam (HSC) on the Subaru Telescope yields the $1\sigma$ SB limit of $(2.5 - 46.0)\times 10^{-19}$ erg s$^{-1}$ cm$^{-2}$ arcsec$^{-2}$ in 1-h exposure \citep{Fab2019,Cai2019,OSullivan2020,Li2024}. 
Given the prediction (Fig.~\ref{probability_stack}), pinpointing the ionization cone requires at least 75 unobscured and obscured AGNs with radio jets detected. 
Although $\geq 300$ type-I Ly$\alpha$ nebulae have been detected at $z\geq 2$ \citep{Borisova2016,Fab2019,Cai2019,Farina2019,OSullivan2020,Mackenzie2021,Fossati2021,Herwig2024,Li2024}, no survey of the Ly$\alpha$ nebulae around obscured AGNs is conducted yet even if there are observations targeting some individual type-II nebulae \citep{Law2018,Brok2020,Zhang2023a, Zhang2023b}. 
We emphasize the need to expand the sample of type-II nebulae around the high-$z$ obscured AGN \citep{alexandroff2013,Wang2025} for revealing the ionization cone to probe the unified model at the $z\geq 2$.


\subsection{The influence of the AGN feedback} \label{open_question}
The AGN feedback is not included in these simulated massive systems in this work \citep{Feldmann2016,Angle2017,Feldmann2017}. 
However, it plays an important role in changing the density, temperature, metallicity, and ionization state of the cool gas reservoirs in the CGM which could influence the properties of nebulae. 
Here, we briefly discuss how the AGN feedback could influence the Ly$\alpha$ nebulae based on the results from previous nebulae simulations. 

Recent simulations predict that the AGN feedback could induce anisotropy in the thermodynamical properties of the CGM gas regardless of the `quasar-mode' feedback where the SMBH has a high accretion rate ($\geq 1\%-10\%$ of Eddington) or the `radio-mode' feedback where the SMBH has low accretion rates \citep{Truong2021,Yang2024}. 
These simulations agree that CGM gas temperature (density) along the axis of the torus will be enhanced (diminished) by 0.1-0.3 dex \citep{Nelson2019,Terrazas2020,Truong2021,Yang2024}. 
According to the empirical relation yielded by \textsc{cloudy} (Sec.~\ref{post_cloudy}), $\leq 0.3$ dex diminishment (enhancement) on $n_{\rm H}$ ($T$) will decrease (increase) the Ly$\alpha$ emissivity by $\leq 0.4$ dex ($\leq 0.2$ dex) for the cool gas ($T= 10^{4}$ K) which dominates the Ly$\alpha$ emission. 
These changes in the CGM thermodynamical properties could lower or raise the Ly$\alpha$ emissivities, i.e. nebulae brightness, depending on the actual status of the CGM gas. 
Whereas, the variation of the nebulae brightness due to the change of gas thermodynamical properties should be minimal (within 0.4 dex).

However, the variation of the gas ionization state due to the AGN feedback could significantly influence the nebulae brightness. 
Simulations with the quasar-mode feedback show that the Ly$\alpha$ nebulae could be strongly boosted by a magnitude of $\approx 10$ if the neutral hydrogen fraction ($X_{\rm HI}$) is suppressed by this feedback mechanism at $z=2-6$ \citep{Byrohl2021,Costa2022}. 
\cite{Byrohl2021} demonstrates that the $X_{\rm HI}$ can be suppressed from $\approx 0.5$ to $\leq 0.2$ for halo gas at $z=3.0$ due to the feedback. 
The suppression on $X_{\rm HI}$ could not only enhance the Ly$\alpha$ emissivity arising from recombination but also increase the escape fraction of the Ly$\alpha$ photons, which makes the nebulae brighter \citep{Costa2022}. 
Considering the anisotropic nature of the quasar-mode feedback under the unified model, the Ly$\alpha$ emissivity of regions out of the ionization cone should be less boosted than regions within the cone. 
This could enhance the anisotropy of the spatial distribution of the Ly$\alpha$ emissivity, which could make observed differences between the two types of Ly$\alpha$ nebulae in the morphology, SB profile, and spectral profile originating from the projection effect of the cone more pronounced.






Unfortunately, how the radio-mode feedback influences the nebulae is unexplored although Sec.~\ref{nebulae_probing_AGN} demonstrates that this feedback mechanism is essential for testing the AGN unified model through nebulae. 
In the future, we will employ simulations with different AGN feedback mechanisms equipped, such as the \textsc{gizmo-simba} run of \textsc{the three hundred} project \citep{Cui2022}, to explore how different AGN feedback mechanisms (quasar-mode or radio-mode) could influence the nebulae and using nebulae to test the AGN unified model.

\section{Conclusions} \label{Conclusions}
In this work, ten massive systems ($M_{\rm h}=10^{12.01-12.46} \ M_{\odot}$) at $z=2-3$ selected from the FIRE cosmological zoom-in simulations \citep{Angle2017,Feldmann2016,Feldmann2017} are used to simulate the Ly$\alpha$, H$\alpha$, and \heii \ nebulae around the unobscured and obscured AGN. 
The ionizing radiation is composed of the anisotropic AGN radiation and the isotropic radiation of the host galaxy and the UVB. 
The AGN intrinsic luminosity and the half-opening angle of the torus are taken as free parameters. 
The gas emissivities are calculated with \textsc{cloudy} \citep{Ferland2017}. 
We project the emissivities of H$\alpha$ and \heii \ to the 2D mesh to generate the mock images. 
As for the Ly$\alpha$, we perform radiative transfer calculations with the \textsc{rascas} code \citep{Michel2020} to generate the mock images and datacubes. 
By analyzing the mock data, we find:

\begin{itemize}
    \item By selecting the emitting region with the dimming-corrected SB threshold which corresponds to $\sigma_{\rm SB}=3.5\times 10^{-18}$ erg s$^{-1}$ cm$^{-2}$ arcsec$^{-2}$ at $z\approx 3.2$, the simulated Ly$\alpha$ nebulae have the luminosity and area ranging in $1.1\times 10^{41} - 3.7\times 10^{44}$ erg s$^{-1}$ and $20 - 5175$ kpc$^{2}$. 
    The type-I and type-II Ly$\alpha$ nebulae do not exhibit significant differences in luminosity and area. 
    Comparing with observations \citep{Fab2019,Cai2019,Fab2023}, our simulations can reproduce the observed luminosity-area relation of $z=2-3$ under the 2-$\sigma$ uncertainty. 
    
    \item The type-I Ly$\alpha$, H$\alpha$, and \heii \ nebulae have higher flux-weighted $\alpha$ parameter ($\alpha_{\rm w}$) than the type-II nebulae, indicating that the type-I nebulae have more symmetric morphologies if the half-opening angle of the torus ($\Delta$) and AGN intrinsic luminosity ($L_{\rm AGN}$) are controlled. 
    The $\alpha_{\rm w}$ is less sensitive to the chosen SB threshold than the flux-unweighted $\alpha$ ($\alpha_{\rm uw}$), which makes $\alpha_{\rm w}$ better at characterizing the nebulae symmetry than $\alpha_{\rm uw}$ in observations. 
    The resonant scattering effect could make nebulae more symmetric ($\alpha_{\rm w}$ is enlarged by $\geq 14\%$) because the Ly$\alpha$ photons are redistributed isotropically through multiple random scattering events. 
    Moreover, our simulations could reproduce the observed symmetry ($\alpha_{\rm w}=0.65\pm 0.19$) of the type-I nebulae at $z=2-3$ under the same SB threshold \citep{Cai2019,Fab2019,OSullivan2020}.
    
    \item The SB profiles of type-I nebulae have smaller scale lengths ($r_{h}$) than the type-II nebulae, indicating that type-I nebulae have steeper profiles than the type-II nebulae if the $\Delta$ and $L_{\rm AGN}$ are controlled. 
    The resonant scattering could also make the Ly$\alpha$ nebulae flatter ($r_{h}$ increases by $\approx 13\%$) by transporting Ly$\alpha$ photons generated at the small radii to the large radii. 
    The comparison between our simulated Ly$\alpha$ nebulae and the observed Ly$\alpha$ nebulae at $z=2-3$ \citep{Fab2019,Cai2019,OSullivan2020,Fossati2021} favors the half-opening angle of torus to be $\Delta =60^{\rm o}$, which is consistent with previous simulations \citep{Obreja2024}. 
    Moreover, after stacking the nebulae by aligning them along the axis of the torus, the type-II nebulae exhibit significant elongation while the type-I nebulae do not. 
    Since the torus is coaxial with the radio jets \citep{Vernet2001,Drouart2012}, the joint observations of nebulae and radio jets provide a way to directly reveal the ionization cone to probe the AGN unified model at high redshift. 
    
    \item The line width (characterized by velocity dispersion, i.e. $\sigma_{v}$) of the spectra of the type-II Ly$\alpha$ nebulae evolves from larger to smaller than that of the type-I Ly$\alpha$ nebulae from the inner to the outer region. 
    Their offset in $\sigma_{v}$ is $\approx 20$ km s$^{-1}$ at $r\geq 10$ kpc, hard to be resolved by current instruments. 
    The resonant scattering effect could increase the $\sigma_{v}$ by $\approx 100$ km s$^{-1}$ through redistributing the Ly$\alpha$ photons from the line center to the wing in the velocity space. 
    Compared with observations \citep{Zhang2023a}, our simulations indicate that the resonant scattering is non-negligible for broadening the line width of the Ly$\alpha$ nebulae. 

    \item {Our simulations can reproduce the observed \heii/Ly$\alpha=0.01-0.1$ \citep{Cai2017,Marino2019,Brok2020,Guo2020,Sanderson2021,Lau2022} and predict that the two types of nebulae should share a consistent radial profile in $10 - 100$ kpc at fixed bolometric luminosity. 
    This result does not favour previous calculations \citep{Cantalupo2019} because volumes out of the AGN's ionization cone also have a significant effect on regulating the \heii/Ly$\alpha$. 
    However, we caution that further tests should be carried out to explore the effects of considering larger simulated volumes (Mpc-scale) and simulations including AGN feedback. 
    Though previous works showed that Ly$\alpha$ and \heii \ nebulae around type-I AGNs can be explained just by using gas distribution on halo scales \citep{Costa2022,Obreja2024}.}
    
    \item The connection between the properties of nebulae and AGN is complicated. 
    The symmetry of the two types of nebulae shows no correlation with $L_{\rm AGN}$. 
    The symmetry of type-I nebulae does not correlate with $\Delta$ while the symmetry of the type-II nebulae correlates with $\Delta$ non-linearly. 
    The $r_{h}$ of all nebulae do not show significant correlation with $\Delta$ while the $r_{\rm h}$ of the Ly$\alpha$ is anti-correlated with $L_{\rm AGN}$. 
    The normalization parameter ($C$) which characterizes the brightness level of all nebulae shows significant negative and positive correlations with the half-opening angle of the torus and the AGN intrinsic luminosity, respectively. 
    These correlations allow us to constrain the AGN engine through nebulae but make testing the AGN unified model impossible by only nebulae observations.

    \item By adopting the method of Sec.~\ref{torus_gal_stacking}, the elongation of the type-II nebulae and non-elongation of the type-I nebulae pinpoint the AGN ionization cone. 
    Our calculations suggest that $\geq 75$ type-II Ly$\alpha$ nebulae with radio jet detected are required to reveal the ionization cone to probe the unified model at a confidence level of 95\% with current instruments. 
    Since current detection of the type-II Ly$\alpha$ nebulae is only limited to a few cases \citep{Law2018,Brok2020,Zhang2023a,Zhang2023b}, we emphasize the need to expand the sample of type-II nebulae for probing the AGN model at the high redshift.

\end{itemize}

Since the simulated systems are not equipped with AGN feedback, we also discuss the influence of AGN feedback on nebulae. 
By adopting the results of previous simulations \citep{Byrohl2021,Costa2022}, we find that, under the unified model, the quasar-mode feedback can boost the nebulae within the ionization cone with a magnitude of $\approx 10$ by suppressing the neutral hydrogen fraction. 
This will enhance the anisotropy of the spatial distribution of the Ly$\alpha$ emissivity. 
The differences between the two types of nebulae in the morphology, SB profile, and spectral profile will become more pronounced. 
Since no simulation of nebulae with radio-mode feedback is included, it is unclear how this feedback mechanism would influence the Ly$\alpha$ emission. 
Our findings suggest that nebulae can help us understand not only the large-scale gas environment of the AGN but also its properties.  
Expanding the joint observations of the type-II Ly$\alpha$ nebulae and radio jets and conducting simulations with different AGN feedback mechanisms are needed to allow us to have a deeper understanding of the galaxy-CGM ecosystem.

\section*{Acknowledgements}

SZ is funded by the China National Postdoctoral Program for Innovative Talents. 
AO's contribution to this project was made possible by funding from the Carl Zeiss Foundation.
\section*{Data Availability} 
The data of the FIRE cosmological simulations is public. 
The codes used in this work are available upon request to the corresponding author. 
The tables generated with \textsc{cloudy} are also available upon request to the corresponding author.



\bibliographystyle{mnras}
\bibliography{example} 




\appendix

\section{Method of producing the mock observations} \label{appendix_mockobs}

\subsection{Modeling the Spectral Energy Distribution (SED) of the ionization source} \label{ionization_sed}

To simulate the CGM nebulae, ionizing photons from the ultraviolet background (UVB), the AGN, its host galaxy, and the satellites should be taken into account. 
In this study, we neglect the ionizing radiation of satellite galaxies. 
Observations have shown that the stellar mass of the satellite to the main galaxy is $< 0.1$ in a halo of $M_{\rm h}=10^{12-13} \ M_{\odot}$ at $z=2$ \citep{Shuntov2022}. 
The satellite fraction in a halo is below 0.2 at this redshift \citep{Shuntov2022}. 
This indicates that the contribution of satellites to induce the nebulae is negligible compared to the host galaxy. 
Moreover, this study focuses on revealing the nebulae induced by the two types of AGN where the radiation contamination from the satellites should be removed. 
In the following subsection, we introduce how we construct the combined SED of the UVB, AGN, and its host galaxy.





\subsubsection{AGN SED under the unified model} \label{agn_sed}
The AGN intrinsic luminosity is set with Eq.~\ref{L_AGN} 
\begin{subequations}
\label{L_AGN}
\begin{align}
    & L_{\rm Edd}= 1.3 \times 10^{46} ({\rm M}_{\rm BH}/10^{8} {\rm M}_{\odot}) \ {\rm erg \ s^{-1}} \\
    & L_{\rm AGN}=\lambda L_{\rm Edd}
\end{align}
\end{subequations}
where $L_{\rm Edd}$ is the Eddington luminosity \citep{Rybicki1986}, and $\lambda$ is the Eddington ratio quantifying the AGN radiative efficiency. 
Since previous observations ($\geq 30000$ sources) have shown that the $\lambda$ centers at $\lambda=0.1 - 0.2$ for broad-line AGN with $M_{\rm BH}=10^{8}-10^{10} \ M_{\odot}$ at the redshift of $2.0\leq z\leq 3.0$ \citep{Shen2011,Suh2015,Vietri2020}, we fix $\lambda=0.1$ in this work. 
Given the mass of the SMBH in Tab.~\ref{halo_info}, we calculate the $L_{\rm AGN}=2.3- 15.2  \times 10^{45}$ erg s$^{-1}$ (Tab.~\ref{skirtor_param}). 
Besides, the shape of the AGN SED is mainly controlled by the viewing angle and properties of the SMBH accretion disc, the torus, and the pole dust \citep{Fritz2006}. 
In this work, we use \textsc{x-cigale} \citep{Yang2020} to build the SED for the two types of AGN under the unified model. 
\textsc{x-cigale} is a multi-wavelength fitting code developed based on \textsc{cigale} \citep{Boquien2019} with the X-ray module included. 
This code has the build-in AGN module, \textsc{skirtor} \citep{Stalevski2012,Stalevski2016}, which is composed of the thermal radiation from the accretion disk \citep{Feltre2012}, the X-ray radiation from the Compton scattering of the hot medium, the infrared (IR) radiation from the polar dust and the dusty torus \citep{Yang2020}. 
In this AGN module, \cite{Stalevski2016} use the clumpy two-phase model based on the 3D radiative-transfer code, \textsc{skirt} \citep{Baes2011,Camps2015}, to produce the dusty torus radiation. 
Since \cite{Yang2020} have successfully used the \textsc{x-cigale} to fit 590 AGNs (206 unobscured AGNs and 384 obscured AGNs) at $z=0.4 - 2.0$ in the COSMOS field, which yields the median $\chi^{2}_{\rm red}$ of 1.4 and 0.9 for unobscured and obscured AGNs, we fix most of the parameters except the L$_{\rm AGN}$, half-opening angle of the torus ($\Delta$), viewing angle ($\theta$) to their results (Tab.~\ref{skirtor_param}). 
The parameters of the \textsc{skirtor} module are listed in Tab.~\ref{skirtor_param}. 
\begin{table*}
    \centering
    \begin{tabular}{cccccccccccc}
    \hline
    \hline
         $\tau_{\rm 9.7}^{(1)}$ & $p^{(2)}$  & $q^{(3)}$ & $\Delta$ [deg]$^{(4)}$ & $R^{(5)}$ & $\theta$ [deg]$^{(6)}$ & frac$_{\rm AGN}^{(7)}$ & $E (B-V)^{(8)}$ & T$_{\rm polar}$ [K]$^{(9)}$ & L$_{\rm AGN}$ [$\times 10^{45}$ erg s$^{-1}$]$^{(10)}$ & $\Gamma^{(11)}$ & $\epsilon_{\rm polar}^{(12)}$\\
         \hline
          7.0 & 1.0 & 1.0 & 50.0 -70.0 & 20.0 &  0.0 - 90.0 & 0.95 & 0.0 & 100.0 & 2.3 - 15.2 & 1.8 & 1.6 \\
         \hline
         \hline
    \end{tabular}
    \caption{{\bf The parameters of the \textsc{skirtor} module.} Col. (1): the torus optical depth at 9.7 $\mu$m. Col. (2): the dusty torus density radial parameter $p$ ($\rho \propto r^{-p} e^{-q|cos\theta|}$. Col. (3): the dusty torus density angular parameter $q$. Col. (4): the half-opening angle of the dusty torus relative to the equatorial plane. Col. (5): the ratio between the outer and inner radii of the dusty torus. Col. (6): the observer's viewing angle relative to the equatorial plane of the torus. Col. (7): the AGN fraction in total IR luminosity. Col. (8): the extinction of the polar dust. Col. (9): the temperature of the polar dust. Col. (10) the AGN intrinsic luminosity derived from Eq.~\ref{L_AGN} by adopting $\lambda=0.1$. Col. (11) the AGN photon index in the X-ray range. 
    Col. (12) the emissivity index of the polar dust. 
    Except for the $\Delta$, $\theta$, and $L_{\rm bol}$, the rest parameters are fixed to best-fit parameters of observations \citep{Yang2020}.}
    \label{skirtor_param}
\end{table*}

For each massive system, the $L_{\rm AGN}$ inputted to the \textsc{x-cigale} is calculated from Eq.~\ref{L_AGN}. 
The viewing angle ($\theta$) is measured as the angle between the sightline and the equatorial plane of the dusty torus. 
We use the default setting of \textsc{x-cigale} to let this value vary in $0^{\rm o}-90^{\rm o}$ in a step of 10$^{\rm o}$. 
For each viewing angle, we assign the corresponding attenuation curve from \cite{Stalevski2016} to the AGN SED of $\theta=90^{\rm o}$ where the AGN is face-on to the observer. 
This process ensures that the obscured and unobscured AGNs share the same intrinsic SED at the UV to X-ray range. 
\cite{Fritz2006} found that the opening angle of the AGN ionization cone should be mostly restricted to $40^{\rm o}\leq \alpha_{\rm cone} \leq 80^{\rm o}$. 
Given $\alpha_{\rm cone}=90^{\rm o}-\Delta$, this range corresponds to the $\Delta$ of $50^{\rm o}\leq \Delta \leq 70^{\rm o}$. 
We, thus, let the $\Delta$ vary in this range in a step of $10^{\rm o}$. 

\begin{figure}
    \centering
    \includegraphics[width=\columnwidth]{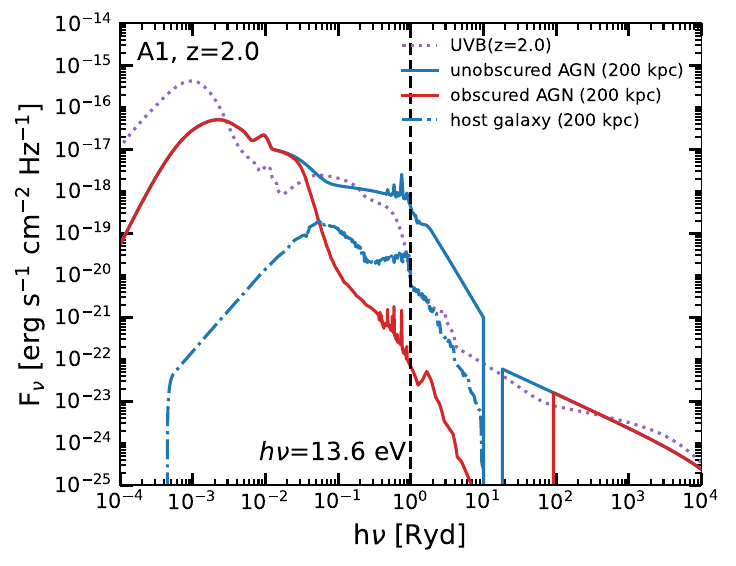}
    \caption{An example of the SED of ionizing radiation including the AGN, its host galaxy, and the UVB. 
    The SED of the AGN and its host galaxy are constructed with \textsc{x-cigale} \citep{Yang2020} based on the $\rm A_{\rm 1}$ system at $z=2.0$ with $M_{\star}=10^{11.0} \ M_{\odot}$ and $M_{\rm BH}\approx10^{8.9} \ M_{\odot}$ (Tab.~\ref{halo_info}). 
    The solid lines denote the SEDs of the unobscured (blue) and obscured (red) AGN, respectively. 
    The dotted-dashed line denotes the SED of its host galaxy and the dotted purple line denotes the UVB SED. 
    The SEDs of the AGN and host galaxy are seen from a distance of 200 kpc. 
    The vertical dashed line denotes the energy of the ionizing photons ($E\approx 13.6$ eV). }
    \label{agn_sed_fig}
\end{figure}

Moreover, \cite{Fossati2021} has shown that the UV line emission of the AGN could influence the observed line ratio on the nebulae. 
For this reason, we add emission lines on top of the AGN continuum. 
We use the composite spectra from \cite{Lusso2018} where the spectra are based on a sample of 104 unobscured AGNs at $z=2.0 - 2.5$ with the $i$ band magnitude of 17.9 - 22.0 and \cite{Mignoli2019} where the spectra are 90 obscured AGNs at $z=1.5 - 3.0$ with FWHM$_{\rm CIV}\leq$ 1350 km s$^{-1}$ for the two types of AGN. 
Fig.~\ref{agn_sed_fig} shows an example of the AGN SED. 
The continuum of the unobscured AGN is $\sim 10^{4}$ higher than that of the obscured AGN at $h\nu = 1$ Ryd.


\subsubsection{Host galaxy SED} \label{galaxy_sed}
Fig.~\ref{agn_sed_fig} shows that the ionizing flux density of the host galaxy could be one order of magnitude higher than that of the obscured AGN. 
This means that the ionizing photons from the host galaxy are not negligible for the region out of the ionization cone. 
Since we do not care about the anisotropy of the galaxy radiation in this work, we simplify the modeling of the gas SED by modeling it as a whole instead of modeling the spectra for each star particle.

We adopt the galaxy model setting of \cite{Yang2020}. 
Specifically, the delayed star formation history (SFH) is employed because it can characterize the SEDs of both early-type and late-type galaxies \citep{Ciesla2015,Boquien2019}. 
To construct the SED, the stellar mass, age, and metallicity are fixed to the stellar mass, mass-weighted age, and mass-weighted metallicity of stellar particles from the FIRE simulations. 
The built-in module \texttt{dustatt\_calzleit} \citep{Calzetti2000,Leitherer2002} of \textsc{cigale} is adopted for modeling the galactic dust attenuation. 
The $E(B-V)$ of the young star is allowed to vary in 0.1 - 0.9 while the $E(B-V)$ ratio between the old and young star is fixed to 0.44. 
We let the amplitude of the 217.5 $\mu$m bump on the extinction curve to be the type of Milky Way. 
We let some parameters vary to ensure that the stellar mass is fixed. 
\begin{table*}
    \centering
    \begin{tabular}{ccccccc}
    \hline
    \hline
         $\tau_{e}$ [Gyr]$^{(1)}$ & $\tau_{s}$ [Gyr]$^{(2)}$ & $Z_{\star}$ [$Z_{\odot}$]$^{(3)}$ & $E(B-V)_{\rm young}^{(4)}$& R$_{E(B-V)}^{(5)}$& $\alpha_{\rm dust}^{(6)}$ & M$_{\star}$ [M$_{\odot}$]\\
         \hline
          0.1, 0.5, 1.0, 5.0& - & - & 0.1 - 0.9& 0.44& 1.5, 2.0, 2.5 & -\\
         \hline
         \hline
    \end{tabular}
    \caption{{\bf The parameters of the module of galaxy SED.} Col. (1): the $e$-folding time of the delayed SFH model. Col. (2): the stellar age of the delayed SFH model. Col. (3): the star metallicity. Col. (4): the $E(B-V)$ of the young population. Col. (5): the $E(B-V)$ ratio between the old and young population. Col. (6) the slope in $dM_{\rm dust} \propto U^{-\alpha} dU$ of the galactic dust re-emission model \citep{Dale2014}. Col. (7): the stellar mass of the host galaxy. 
    The ``-'' denotes that this parameter is set to the value from the FIRE simulations.}
    \label{galaxy_param}
\end{table*}
Fig.~\ref{agn_sed_fig} shows an example of the SED of the host galaxy. 
The galaxy SED is $\approx 100$ times lower than that of the unobscured AGN at $h\nu\geq 1$ Ryd. 
By adopting the galaxy escape fraction from \citep{Khaire2019} which yields $f_{\rm esc}\approx 10^{-3} - 10^{-2}$ at $z=2-3$, the ionizing photon flux from the galaxy should be $10^{4} - 10^{5}$ times lower than that of the unobscured AGN but comparable to that of the obscured AGN and the UVB.

\subsubsection{UVB SED} \label{uvb_sed}
Since the flux of the ionizing photons from UVB is comparable to the host galaxy and the obscured AGN, ionizing photons from UVB are not negligible for regions out of the ionization cone. 
We employ the results of \cite{Khaire2019} to model the UVB SED at $z=2-3$ in this work (Fig.~\ref{agn_sed_fig}).







\subsection{Processing with \textsc{cloudy}} \label{post_cloudy}

To simulate the nebulae of Ly$\alpha$ 1215, H$\alpha$ 6563, and  \heii \ 1640, we run a large grid of photoionization model with the \textsc{cloudy} code \citep{Ferland2017}. 
The cloud is fixed to the plane-parallel geometry. 
The parameter setting for \textsc{cloudy} is the following: (i) ${\rm log} (n_{\rm H}$/cm$^{-3}) = -5 - 2$ in a step of 1 dex, (ii) ${\rm log}( Z/Z_{\odot}) = -4 - 1$ in a step of 1 dex, and (iii) ${\rm log} (T/K) = 3 - 7$ in a step of 1 dex. 
(iv) For the surface flux of the ionizing photon ($\Phi(H)$), the contribution of the UVB is fixed to $\Phi(H)_{\rm UVB}=10^{5.67} \ {\rm cm^{-2} \ s^{-1}}$ which is calculated from the UVB SED. 
The contribution of the AGN and host galaxy at different radii is calculated by Eq.~\ref{surface_flux_ionizing_photon} 
\begin{subequations}
\label{surface_flux_ionizing_photon}
\begin{align}
    & \Phi(H)_{\rm AGN+host}=\frac{Q(H)}{4\pi r^{2}} \\
    & Q(H)=\int_{\nu_{0}}^{\infty}\frac{f_{\nu}d\nu}{h\nu}
\end{align}
\end{subequations}
where $Q(H)$ is the number of the ionizing photons calculated from the SED of the AGN and its host and $\nu_{0}\approx 912$ \AA \ is the Lyman-limit frequency. 
We let the radius vary in $1-200$ kpc in a step of $\approx 17$ kpc (13 bins). 
Eq.~\ref{surface_flux_ionizing_photon} is based on the assumption that the CGM is fully ionized. 
This assumption is not entirely correct for the whole domain (correct for regions within the cone but incorrect for regions out of the cone, see Appendix~\ref{phi_approximation}). 
However, adopting this assumption does not weaken the conclusion (Appendix~\ref{phi_approximation}). 

Considering the variation of viewing angle ($\theta$) and the half-opening angle of the torus ($\Delta$), the whole parameter grid yields 29250 models for each system. 
We do not include molecules, dust, and cosmic rays because the simulations do not follow them. 
For the self-shielded gas ($T\leq 10^{5}$ K), we adopt the column density criteria of \cite{Ploeckinger2020} (Eq.~\ref{columnN_cloudy})
\begin{subequations}
\label{columnN_cloudy}
    \begin{align}
        &  {\rm log}(N_{\rm criteria})={\rm log}(N_{c})-\frac{{\rm log}(N_{c})-{\rm log}(N_{\rm min})}{1+(\sqrt{T_{\rm min}T_{\rm max}}/T)^{k}}\\
        & {\rm log}(N_{c})={\rm min}[{\rm log}(N_{\rm J}), {\rm log}(l_{\rm max}n_{\rm H}), {\rm log}(N_{\rm max})] \\
        & {\rm log}(N_{J})=19.44+0.5[{\rm log}(n_{\rm H})+{\rm log}(T)]
    \end{align}
\end{subequations}
where ${\rm log}(N_{J})$ is the Jeans column density \citep{Schaye2001}. 
The $T_{\rm max}$, $T_{\rm min}$, $l_{\rm max}$, $N_{\rm max}$ and $k$ are set to $T_{\rm max}=10^{5}$ K, $T_{\rm min}=10^{3}$ K, $l_{\rm max}=100$ kpc, $N_{\rm max}=10^{24} \ {\rm cm^{-2}}$, and $k\approx 2.17$, respectively by following \cite{Ploeckinger2020}. 
The $N_{\rm min}$ is set to $N_{\rm min}=l_{\rm max}n_{\rm H, min}\approx 3.1 \times 10^{15} \ {\rm cm^{-2}}$ where the $n_{\rm H, min}=10^{-8} \ {\rm cm^{-2}}$ \cite{Ploeckinger2020}. 
For the unshielded situation ($T> 10^{5}$ K), we let the \textsc{cloudy} code stop after the first zone where the maximum depth of the first zone is set to $\Delta r=10^{20} \ {\rm cm}$ \citep{Ploeckinger2020}. 
This stop column density is based on the self-gravitating gas \citep{Schaye2001}. 
The above settings avoid the column density being too high to induce the practical problems in \textsc{cloudy} or too low to lead to an unrealistic length scale. 
Please find the detailed explanation of the stop column density in \cite{Ploeckinger2020} (Sec.~2.1). 
The outputs of these photoionization models include the heating and cooling rate, the electron density, the ionization fraction for a few species, and the emissivities of different lines. 
All these outputs are saved for the last \textsc{cloudy} zone. 
For gas particles within a grid cell of the \textsc{cloudy} parameter space, the emissivities of different lines are assigned to them by linear interpolation.


\subsection{Simulate the nebulae} \label{mock_diffuse_emission}
In this section, we present the method to simulate the nebulae. 
For the non-resonant lines such as H$\alpha$ and \heii, we follow \cite{Obreja2024} to simulate the nebulae.
For the Ly$\alpha$, we further run the radiative transfer (RT) simulations. 
Since the distribution and abundance of the \civ \ 1548, 1550 \ highly correlate with the feedback mechanism equipped in the cosmological simulations, we ignore the \civ \ in this study. 

\subsubsection{Non-resonant lines (H$\alpha$ and \heii)} \label{non_resonant_line}
The emergent emissivity ($\epsilon_{\rm \nu, em}$) of an emission line with the frequency of $\nu$ for a gas particle follows Eq.~\ref{ep_em2in_eq} 
\begin{subequations}
\label{ep_em2in_eq}
    \begin{align}
        & \epsilon_{\rm \nu, em}=\epsilon_{\rm \nu, in} e^{-\tau_{\nu}} \\
        & \tau_{\nu}=n_{\nu}d_{\rm zone}\sigma_{\nu}
    \end{align}
\end{subequations}
where $\epsilon_{\rm \nu, in}$ is the emissivity yielded by \textsc{cloudy}, $\tau_{\nu}$ is the local optical depth at the frequency of $\nu$, $n_{\nu}$ is the number density of the species emitting the emission, $\sigma_{\nu}$ is the absorption cross-section of the emission line, and $d$ is the depth of the last zone. 


For the H$\alpha$ line, we have $\tau_{\rm H\alpha}=\sigma_{\rm H\alpha}n_{\rm HI,2p}d_{\rm zone}$ where $n_{\rm HI, 2p}$ represents the number density of the neutral hydrogen at the first excited state. 
Since the fraction of neutral hydrogen atoms at the first excited state is $\approx 0$, the attenuation of the neutral hydrogen in the CGM on the H$\alpha$ can be neglected. 
For the \heii \ line, the attenuation of the He$^{+}$ in the CGM is negligible for a similar reason as above. 
The dust attenuation on these lines is also negligible since its density is minimal.


After deriving the emergent emissivities, the luminosities of each gas particle can be calculated through Eq.~\ref{Lumin}
\begin{equation}
    L_{\nu, i}=\epsilon_{\nu, em, i}\frac{m_{i}}{\rho_{i}}
    \label{Lumin}
\end{equation}
where $m_{i}$ and $\rho_{i}$ denote the mass and density of the $i$th gas particle. 
Then, the observed flux is calculated as $F_{\nu}=L_{\nu}/4\pi D_{\rm L}(z)^{2}$ where $D_{\rm L}(z)$ is the luminosity distance at the redshift of $z$.  

\subsubsection{Resonant scattering of the Ly$\alpha$} \label{lya_radiative_transfer}
Eq.~\ref{Lumin} is also adopted to calculate the observed flux of the Ly$\alpha$ where $\epsilon_{\nu, Ly\alpha}$ represent the emissivity yielded by \textsc{cloudy}. 
Since we add the line emissions to the AGN SED and allow the continuum radiative pumping in the \textsc{cloudy} calculation, the photons from the BLR of the AGN have been included. 
By observing a type-II nebulae at $z\approx 3.2$, \cite{Sanderson2021} has shown that the resonant scattering of Ly$\alpha$ could play an important role in regulating the morphology and kinematics of the nebula. 
Thus, we adopt the \textsc{rascas} code \citep{Michel2020} to simulate the scattering of the Ly$\alpha$ photons in this work. 
 
\textsc{rascas} is a public three-dimensional (3D) radiative transfer code that performs resonant photon propagation on an adaptive mesh in the octree structure.  
Since FIRE simulations are run with the \textsc{gizmo} code based on the SPH method (Sec. \ref{simulations}), we utilize the python module \textsc{yt}\footnote{\url{https://yt-project.org/}} \citep{Turk2011} to project gas particles to an AMR. 
To balance the computing time and accuracy, we refine AMR cells only when they contain $\geq 15$ particles. 
This yields the smallest cell size to be 0.05 kpc. 
The $n_{\rm H}$ and $n_{\rm HI}$ are assigned as $n_{\rm j}=\Sigma_{i}n_{\rm i,j}V_{\rm i,j}/V_{\rm j}$ where $n_{\rm j}$ denote the $n_{\rm H}$ ($n_{\rm HI}$) in the $j$th leave cell, $V_{\rm j}$ is the volume of this leave cell, $n_{\rm H,i,j}$ is the hydrogen number density of the $i$th gas particle in this cell, and $V_{\rm i,j}$ is the volume of the $i$th gas particle. 
This ensures that the total number of hydrogen atoms in the cell is conserved. 
The temperature is assigned as the density-weighted temperature, and the velocity is assigned as the HI-mass-weighted velocity. 
For the Ly$\alpha$ emissivity, we assign each cell with the volume-weighted emissivity. 
This ensures that the total luminosity does not change after projecting gas particles on the mesh.


450000 photon packets are adopted as we find that the surface brightness (SB) profile converges under this number of photon packets.
We assign the photon packets to the mesh cell where the number of photon packets is proportional to the cell luminosity. 
For the frequency of the photon packet, it is first randomly drawn in the reference frame of the cell by assuming a Gaussian line profile centered on the frequency of $\nu_{Ly\alpha, 0}=2.47\times 10^{15}$ Hz. 
The frequency is then shifted to the external frame according to the cell's velocity under the Doppler effect. 
For the direction of the photon packet, we initialize it with a random orientation. 
In each scattering event, the outcoming direction is related to the incoming direction and the frequency of the photon packet in the scatter's frame through the phase function introduced in \cite{Michel2020}.

We adopt the build-in method of \textsc{rascas} to model the dust absorption of the Ly$\alpha$ emission \citep{Laursen2009}. 
Under this model, the dust number density correlates with the gas metallicity, hydrogen density, and the fraction of ions ($f_{\rm ion}$), which is calculated as Eq.~\ref{dust_number_density}
\begin{equation}
    n_{\rm dust}=\frac{Z}{Z_{0}}(n_{\rm HI}+f_{\rm ion}n_{\rm HII})
    \label{dust_number_density}
\end{equation}
where $f_{\rm ion}=0.01$, $Z_{0}=0.01 (0.005)$ is the mean metallicity of the Small Magellanic Cloud (Large Magellanic Cloud), and $n_{\rm HII}$ is the number density of the proton. 
Since the dust absorption cross-section ($\sigma_{\rm dust}$) is largely independent of the frequency \citep{Laursen2009}, we adopt the model of `Small Magellanic Cloud (SMC)' with the constant cross-section of $\sigma_{\rm dust}/m_{\rm p}\approx 960$ cm$^{2}$ g$^{-1}$ in \textsc{rascas} \citep{Michel2020,Costa2022}. 
As for the scattering of the Ly$\alpha$ photons, the \textsc{rascas} code adopts the phase function of \cite{Laursen2009} to describe this probability.

\subsection{Constructing the mock observables} \label{mockobs}

Before reproducing the observables of the nebulae, it is necessary to clarify the setting for generating the mock observables. 
As mentioned in Sec.~\ref{agn_sed}, we let the half-opening angle of the torus ($\Delta$) to vary in $50^{\rm o}-70^{\rm o}$ in a step of $10^{\rm o}$. 
For each $\Delta$, we randomly select three orientations to place the ionization cone.
For each orientation of the cone, we randomly place a sightline inside and outside the cone to mock the observables for the two types of AGN, respectively. 
We then construct the mock images for the Ly$\alpha$ without the processing of \textsc{rascas}, Ly$\alpha$ with the processing of \textsc{rascas}, H$\alpha$, and \heii. 
These settings yield $10\times 3\times 3\times 2=180$ mock observables (images and datacubes) for each emission line where the left-side numbers of the equation denote the number of systems, placements of the cone, settings of $\Delta$, and types of AGN, respectively.

For the Ly$\alpha$ without the processing of \textsc{rascas}, H$\alpha$, and \heii, we directly project the domain onto a 2D mesh with $200\times 200$ pixels. 
The same image size is adopted for Ly$\alpha$ with the processing of \textsc{rascas} in generating the mock image with \textsc{rascas}. 
These images are then convolved with a 2D Gaussian kernel that has the $\rm FWHM=1''$ to mock the usual seeing in ground-based observations. 
To construct the datacubes for the Ly$\alpha$ emission, the \textsc{rascas} adopt the `peeling algorithm' \citep{Michel2020,Costa2022}. 
We fix the size of the datacube to be $200\times 200\times 100$ where the first two values represent the pixel number in the two spatial directions and the last value represents the number of spectral bins. 
With the spectral range of $-1500 {\rm \ km \ s^{-1}}\leq \Delta v \leq 1500 \ {\rm km \ s^{-1}}$. 
A 1D Gaussian kernel with $\rm FWHM=75 \ {\rm km \ s^{-1}}$ (corresponding to the resolving power of $R=4000$) is convolved to the spectra of each voxel to simulate the line spread function (LSF) of the BM grating of the Keck/KCWI \citep{Morrissey2018}.






\section{Estimate of the surface flux of the ionizing photons ($\Phi(H)$)} \label{phi_approximation}

In this section, we estimate the radial profile of $\Phi(H)$ when the absorption of the neutral hydrogen and dust in the CGM is taken into account. 
We note that an accurate radial profile of $\Phi(H)$ requires radiative transfer simulations. 
Whereas, the analytical estimate can help us know the upper and lower limits of the $\Phi(H)$.

The $Q(H)$ at the radius of $r$ should follow Eq.~\ref{QH_absorption} if the absorption of the neutral hydrogen and the dust on the ionizing photons is included 
\begin{subequations}
\label{QH_absorption}
\begin{align}
    & Q(H,r)=\int_{\nu_{0}}^{\infty} \frac{f_{\nu}e^{-\tau_{\nu}(r)}}{h\nu}d\nu \\
    & \tau_{\nu}(r)=\tau_{\nu, {\rm HI}}(r)+\tau_{\nu, {\rm dust}} (r)
\end{align}
\end{subequations}
where $\tau_{\nu, {\rm HI}}(r)=n_{\rm HI}\sigma_{\nu, {\rm HI}}r$ and $\tau_{\nu, {\rm dust}}(r)=n_ {\rm dust}\sigma_{\nu, {\rm dust}}r$ represent the optical depth of the neutral hydrogen and dust. 
The $\sigma_{\nu, {\rm HI}}$ and $\sigma_{\nu, {\rm dust}}$ represent the absorption cross section of the neutral hydrogen and dust, respectively. 
We adopt the Eq.~\ref{cross_section_HI} from \cite{Osterbrock2006} to calculate the cross section of the neutral hydrogen
\begin{equation}
    \sigma_{\nu, {\rm HI}}=\{^{6.3\times 10^{-18} (\nu_{0}/\nu)^{3} \ {\rm cm^{2}} (\nu\leq \nu_{0})}_{0 \ {\rm cm^{2}} (\nu > \nu_{0})}
    \label{cross_section_HI}
\end{equation}
where $\nu_{0}\approx3.3\times 10^{15} \ {\rm GHz}$ is the Lyman-limit frequency. 
The fit of \cite{Gnedin2008} is adopted to calculate the $\sigma_{\nu,{\rm dust}}$ \citep{Laursen2009,Michel2020}. 
The $n_{\rm dust}$ is calculated with Eq.~\ref{dust_number_density}. 
Since we run the \textsc{cloudy} code under the assumption that the gas is optically thin, the code yields the lower limit of the neutral hydrogen fraction ($x_{\rm HI}$). 
The $x_{\rm HI}$ yielded by only the UVB \citep{Rahmati2013} is adopted as the upper limit. 
The number density of the neutral hydrogen is calculated by $n_{\rm HI}=x_{\rm HI}n_{\rm H}$. 
The escape fraction of the ionizing photons ($f^{\rm ion}_{\rm esc}$) is adopted to quantify the attenuation where  $f^{\rm ion}_{\rm esc}(r)=\Phi_{\rm att}(H,r)/\Phi_{0}(H,r)$. 
The $\Phi_{\rm att}(H)$ is the surface flux after the attenuation and $\Phi_{0}(H,r)$ is the surface flux with no attenuation.  
The radial profile of the upper and lower limits of $f^{\rm ion}_{\rm esc}$ which corresponds to the two limits of $X_{\rm HI}$ is calculated by combining Eq.~\ref{surface_flux_ionizing_photon} and Eq.~\ref{QH_absorption}. 

\begin{figure*}
    \centering
    \includegraphics[width=\textwidth]{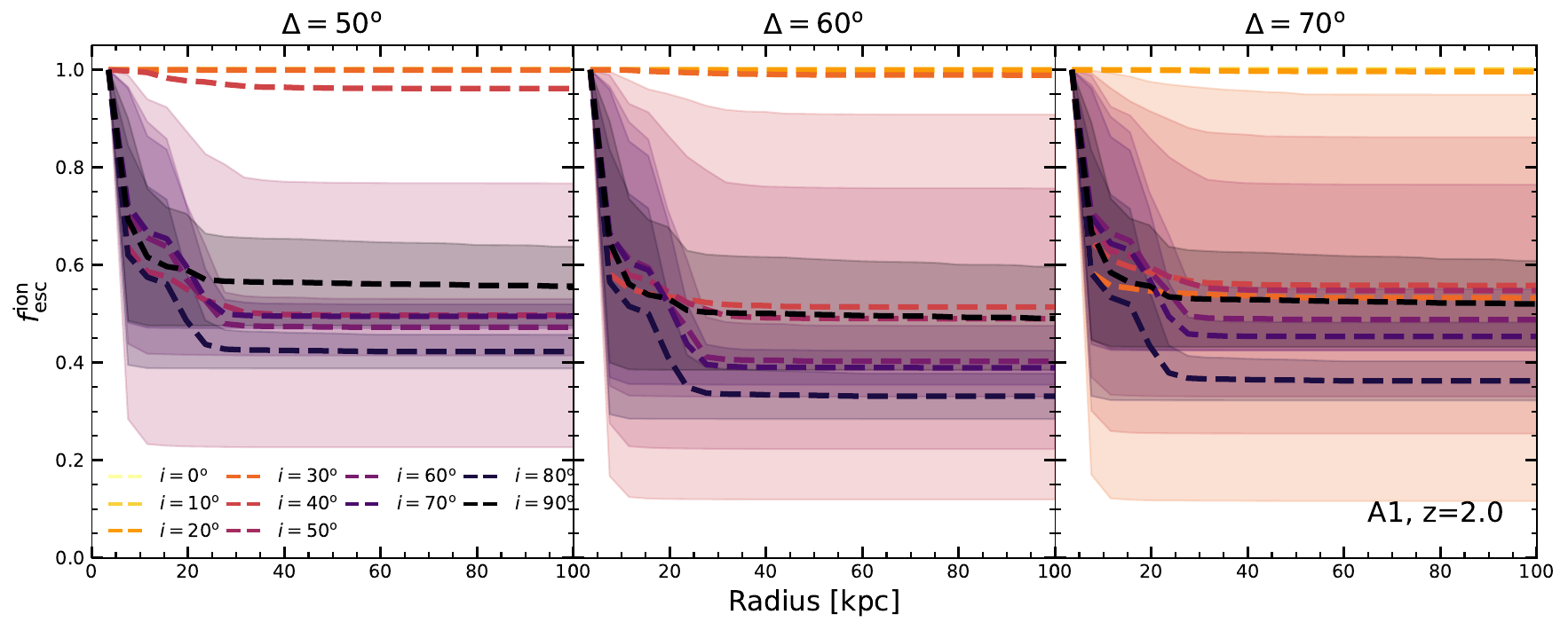}
    \caption{The radial profile of the escape fraction of the ionizing photons for half-opening angles of the torus varying in $50^{\rm o}-70^{\rm o}$. 
    The color of the dashed line denotes the inclination relative to the equatorial plane of the torus. 
    The dashed lines of $\theta> 90^{\rm o}-\Delta$ are the upper limit (applying the $X_{\rm HI}$ from \textsc{cloudy}). 
    The dashed lines of $\theta\leq 90^{\rm o}-\Delta$ are derived as the mean of the upper and lower limits. 
    For $\theta> 90^{\rm o}-\Delta$ (regions within the cone), the $f_{\rm esc}^{\rm ion}$ drops by $\Delta f_{\rm esc}^{\rm ion}\leq 0.05$ indicating that these regions should be fully ionized. 
    For $\theta\leq 90^{\rm o}-\Delta$ (regions out of the cone), the $f_{\rm esc}^{\rm ion}$ drops to $\approx 0.5$ indicating that these regions are not fully ionized. 
    However, we still adopt the assumption that the gas is fully ionized in this work because it does not weaken the conclusion.}
    \label{f_esc_A1}
\end{figure*}

For regions within the ionization cone ($\theta> 90^{\rm o}-\Delta$), we only use the lower limit of $X_{\rm HI}$, i.e. the $X_{\rm HI}$ produced by the \textsc{cloudy}, to approximate the attenuation because the $\Phi(H)$ of the AGN is $\geq 10^{3}$ times larger than that of its host galaxy and the UVB at $r\leq 100$ kpc. 
$f^{\rm ion}_{\rm esc}$ of the $\rm A_{1}$ system at $z=2.0$ is shown in Fig.~\ref{f_esc_A1}. 
The $f_{\rm esc}^{\rm ion}$ of these regions drops by $\Delta f_{\rm esc}^{\rm ion}\leq 0.05$ at $r\leq 100$ kpc, indicating that the assumption of fully ionizing CGM should hold for these regions. 
This result is consistent with\cite{Obreja2024} and \cite{Byrohl2021}. 
For regions out of the cone ($\theta\leq 90^{\rm o}-\Delta$), we take the mean of the upper and lower limits of the $f_{\rm esc}^{\rm ion}$ as the approximation of the real $f_{\rm esc}^{\rm ion}$. 
Fig.~\ref{f_esc_A1} shows that the $f_{\rm esc}^{\rm ion}$ rapidly drops to $f_{\rm esc}^{\rm ion}\approx 0.5$ in $r\leq 20$ kpc. 
According to the relation between $\epsilon_{\rm Ly\alpha}$ ($X_{\rm HI}$) and $\Phi(H)$ produced by \textsc{cloudy}, the Ly$\alpha$ emissivity will be about two times lower (higher) if the $\Phi(H)$ drops by a factor of $\approx 0.5$.

The estimates above indicate that the difference in the brightness of regions within and out of the cone should be even larger if the $\Phi(H)_{\rm att}$ is adopted. 
This will lead to the differences in the morphologies, SB profiles, and spectral profiles of nebulae around the two types of AGN (Sec.~\ref{sim_results}) becoming more significant because these differences originate from the projection effect of the cone. 
Since calculating the $\Phi_{\rm att} (H)$ by the radiative transfer modeling is time-consuming, and applying it to the simulations will strengthen the conclusion instead of weaken it, we adopt Eq.~\ref{surface_flux_ionizing_photon} to calculate the $\Phi(H)$ for regions out of the cone. 


\bsp	
\label{lastpage}
\end{document}